\documentclass[twocolumn,showpacs,prb,amsfonts,amsmath,amssymb,floatfix]{revtex4}
\usepackage{color}
\usepackage{hhline}
\usepackage{mathrsfs}
\usepackage{dcolumn}
\usepackage{multirow}
\usepackage{bm}
\usepackage{afterpage}
\usepackage{amsmath}
\usepackage{booktabs}
\usepackage{graphics}
\usepackage{graphicx}

\newcommand{\ph}{\phantom{0}}
\begin{document}

\title{
High-temperature Superconductivity in Layered Nitrides $\beta$-Li$_{x}$MNCl (M = Ti, Zr, Hf): Insights from Density-functional Theory for Superconductors}

\author{Ryosuke Akashi$^{1}$}
\author{Kazuma Nakamura$^{1,2}$}
\author{Ryotaro Arita$^{1,2,3}$}
\author{Masatoshi Imada$^{1,2}$}
\affiliation{$^1$Department of Applied Physics, The University of Tokyo, 
             Hongo, Bunkyo-ku, Tokyo 113-8656, Japan}
\affiliation{$^2$JST-CREST, Hongo, Bunkyo-ku, Tokyo 113-8656, Japan}
\affiliation{$^3$JST-PRESTO, Kawaguchi, Saitama 332-0012, Japan}

\date{\today}
\begin{abstract}
We present an {\em ab initio} analysis with density functional theory for superconductors (SCDFT) to understand the superconducting mechanism of doped layered nitrides $\beta$-Li$_{x}$MNCl (M=Ti, Zr, and Hf). The current version of SCDFT is based on the Migdal-Eliashberg theory and has been shown to reproduce accurately experimental superconducting-transition temperatures $T_{{\rm c}}$ of a wide range of phonon-mediated superconductors. In the present case, however, our calculated $T_{\rm c}$$\leq$4.3 K (M=Zr) and   $\leq$10.5 K (M=Hf) are found to be less than a half of the experimental $T_{\rm c}$. In addition, $T_{{\rm c}}$ obtained in the present calculation increases with the doping concentration $x$, opposite to that observed in the experiment. Our results indicate that we need to consider some elements missing in the present SCDFT based on the Migdal-Eliashberg theory.
\end{abstract}
\pacs{74.20.Pq, 74.25.Kc, 74.62.-c, 74.70.Dd}

\maketitle

\section{INTRODUCTION}\label{sec:intro}
Since the discovery of high-transition-temperature ($T_c$) superconductivity in cuprates,\cite{Bednorz-Muller} superconductivity in layered transition-metal compounds has been extensively studied. So far, a variety of exotic superconductors such as SrRu$_2$O$_4$, Na$_x$CoO$_2$, iron arsenides and chalcogenides have been found. Among them, layered metallonitride halides (MNX, M=Ti, Zr, Hf; X=Cl, Br, I) discovered by Yamanaka {\it et al.}\cite{Yamanaka1996} are of great interest, for which $T_c$ was found to be as high as 26 K at maximum.\cite{Yamanaka-nat} In fact, they had been the second highest record among transition metal compounds until the recent discovery of iron-based superconductors. The mother compounds of MNX consist of stacking of two-dimensional MN block layers, sandwiched by interlayer halogen ions. Since the interlayer bonding is relatively weak, we can intercalate various Louis bases between the layers or deintercalate halogen atoms for the carrier doping.\cite{Yamanaka-review} Interestingly, while the mother compounds are non-magnetic band insulators, they become superconductors upon doping. This is in high contrast with many other superconducting layered transition metal compounds, for which various magnetic phases reside in the vicinity of the superconducting phase.

When we consider the mechanism of superconductivity in the doped MNX, we have two possibilities. One is a conventional scenario based on the Migdal-Eliashberg (ME) theory,~\cite{Migdal,Eliashberg,Scalapino,Schrieffer} where the superconductivity is mediated by phonons, and the damping and retardation effects are treated by self-consistent perturbation theory, retaining the lowest-order dressed-phonon and dressed-Coulomb contributions to the self energy. The momentum dependences of the gap function, self energy, and screened Coulomb interaction is neglected within the energy scale of phonons. 
In the ME theory, the screened Coulomb interaction is conventionally addressed with the random-phase approximation (RPA) and the frequency dependence is totally ignored. The other scenario is to consider the factors neglected in the ME theory, in which we consider various pairing glues other than phonons, such as spin, charge, or orbital fluctuations. In fact, for some layered transition metal superconductors, it has been believed that the pairing mechanism is unconventional. For example, the anisotropic $d$-wave pairing in cuprates and triplet pairing in Sr$_2$RuO$_4$ are difficult to understand within the ME theory. As in the case of iron-based superconductors, when the electron-phonon coupling is weak,\cite{Boeri2008} we can also safely exclude the possibility of the conventional scenario. On the other hand, the situation is not so simple for doped MNX. 

Experimentally, the Knight shift decreases toward zero below $T_c$, indicating that the Cooper pair is spin singlet.\cite{Tou-isotope, Tou-NMR2010} The break-junction and scanning tunneling spectroscopy measurements suggest that the gap function does not have nodes.\cite{Ekino-tunnel, Takasaki-tunnel2005, Takasaki-tunnel2010} While these are typical behaviors of conventional superconductors, there are also several observations which support unconventional scenario. Measurements of uniform spin susceptibility\cite{Tou-suscep} and specific heat\cite{Taguchi-specheat} indicate that the electronic density of states (DOS) at the Fermi energy ($E_F$) is small and the estimated total electron phonon coupling $\lambda$ is smaller than 0.22. On the other hand, the specific-heat~\cite{Taguchi-specheat} and tunneling-current measurements\cite{Ekino-tunnel, Takasaki-tunnel2005,Takasaki-tunnel2010} give a rather large superconducting gap ratio $\frac{2\Delta}{k_{\rm B}T}$$\simeq$4.6-5.2 or even larger, implying the strong effective pairing interaction between electrons.  
Also, the nitrogen isotope effect coefficient has been reported to be markedly small as $\sim$0.07,\cite{Tou-isotope,Taguchi-isotope} compared with the value of 0.5 for the BCS superconductor.\cite{BCS}
Doping dependence of $T_c$ (Ref.~\onlinecite{Taguchi-doping, Takano-molecular, Kasahara2009}) is also difficult to understand in terms of the conventional scenario; particularly in $\beta$-Li$_{x}$ZrNCl, it becomes higher with decreasing doping and exhibits the highest value at the border of the Anderson insulating phase and the superconducting phase. On top of these, more recently, it has been reported that the spin-lattice relaxation rate of the nuclear-magnetic resonance does not exhibit the coherence peak,\cite{Tou-NMR2007,Kotegawa-JPS2011} and muon spin rotation measurement indicates that the gap function has a significant anisotropy.\cite{Hiraishi}

As for theoretical studies, both the phonon and electron mechanisms have been examined and proposed. For the electron mechanism, 
a possibility of spin-fluctuation-mediated superconductivity has been studied within the lattice Fermion model.\cite{Kuroki2008,Kuroki2010,Kasahara2009, Choi-Kuroki2011, Yin-Pickett} For the phonon mechanism, {\em ab initio} analyses based on the ME theory have widely been performed: Heid and Bohnen~\cite{Heid-DFPT} calculated the phonon spectrum and the electron-phonon coupling using density functional perturbation theory and estimated $T_c$ using the dirty-limit gap equation.\cite{Bergman-Rainer} They found that $\lambda$ is as large as 0.5 for $\beta$-Li$_{1/6}$ZrNCl and the value is too small to explain the experimental value, $T_c$$\sim$16 K, with setting the Coulomb pseudo potential $\mu^{\ast}$ (Ref.~\onlinecite{Morel-Anderson}) to a typical value of $\sim$0.10. Weht \textit{et al.}\cite{Weht-band} evaluated the Hopfield-McMillan parameter $\eta$, also a measure of the electron lattice coupling, for $\beta$-Na$_{0.25}$HfNCl using the Gaspari-Gyorffy formula.\cite{Gaspari-Gyorffy} They found that $\eta_{\rm Hf}$$\simeq$0.5 ${\rm eV/\AA^{2}}$ and $\eta_{\rm N}$$\simeq$0.2 ${\rm eV/\AA^{2}}$, which are one order of magnitude smaller than those of the binary transition-metal-nitride superconductors with $T_{\rm c}$$\simeq$20 K.\cite{Klein-Pickett} 
On the other hand, recently, Yin {\it et al.}\cite{Yin-Kotliar} proposed that $\lambda$ is estimated to be larger if we employ the hybrid Heyd-Scuseria-Ernzerhof functional\cite{HSE} and $T_c$ given by the Macmillan-Allen-Dines (MAD) formula\cite{McMillan, AllenDynes} with $\mu^{\ast}$=0.1 shows a nice agreement with experiments. It should be noted here that there are many examples where $\mu^{\ast}$ substantially (larger than 50\%) deviates from the typical value $\sim 0.1$,\cite{Carbotte} so the problem whether the experimental $T_{\rm c}$ can be described within the ME theory is yet to be clear. We also note that there have been several studies which try to consider the physics beyond the ME theory; for example, Bill {\it et al.} showed that acoustic plasmon can enhance the phonon-mediated superconductivity.\cite{Bill-rapid2002} There are also several studies which propose that disorder can enhance the pairing instability in systems with phonon-mediated attractive interactions.\cite{Kravtsov, Burmistrov}

For the electronic interaction, there are many efforts for describing it from first principles. Lee \textit{et al.}\cite{Lee-Chang-Cohen} proposed a method to evaluate the bare Coulomb parameter $\mu$ with {\em ab initio} RPA calculations. However, there still remains an empirical parameter representing the electronic energy scale, needed for obtaining the renormalized value $\mu^{*}$. On the other hand, L\"uders \textit{et al.} recently developed a parameter-free method to calculate $T_{\rm c}$ based on density functional theory for superconductors (SCDFT),\cite{Luders,Oliveira} where electron-phonon coupling and electron-electron interactions are nonempirically treated. Since the exchange-correlation kernels are constructed on the basis of the ME theory, the current-stage SCDFT has a close correspondence with the ME theory. This method has been applied to a wide range of typical phonon-mediated superconductors such as simple metals,\cite{Marques-metals} MgB$_{2}$,\cite{Floris-MgB2} calcium-intercalated graphite,\cite{Sanna-CaC6} Li, K and Al in high pressure\cite{Profeta-pressure} and CaBeSi,\cite{Bersier-CaBeSi} and the reliability on the quantitative aspect has been examined and established, where $T_c$s have shown agreements with experimental results within a range of few K.

In this work, we performed an SCDFT analysis for the lithium-doped $\beta$-MNCl (M=Ti, Zr, Hf) superconductors to examine whether they can be described by the current SCDFT formalism based on the ME theory. We calculate electronic structures, phonon spectra, electron-phonon couplings, screened Coulomb interactions in the RPA level to estimate superconducting $T_{\rm c}$ and examine the results focusing on the metallic-atom and doping dependences. We found that $T_{\rm c}$ estimated by SCDFT is at maximum half of the experimental $T_c$ and its doping dependence is opposite to experiments, both of which imply that, as in the unconventional superconductors, the present superconductors require some elements missing in the conventional ME theory.

In Sec.~\ref{sec:method}, we review the SCDFT formalism\cite{Luders, Marques-metals, Massidda} and describe our computational detail. Applications to aluminum and niobium are performed in Sec.~\ref{sec:metal} to test the reliability of our calculations. We show results for $\beta$-Li$_x$MNCl in Sec.~\ref{sec:result}. Section~\ref{sec:discussion} is devoted to discussions. Summary and outlook are drawn in Sec.~\ref{sec:sum}. 

\section{METHOD}\label{sec:method}
In SCDFT, we solve the following gap equation\cite{Luders, Marques-metals}
\begin{eqnarray}
\Delta_{n{\bf k}}\!=\!-\mathcal{Z}_{n\!{\bf k}}\!\Delta_{n\!{\bf k}}
\!-\!\frac{1}{2}\!\sum_{n'\!{\bf k'}}\!\mathcal{K}_{n\!{\bf k}\!n'{\bf k}'}
\!\frac{\mathrm{tanh}[(\!\beta/2\!)\!E_{n'{\bf k'}}\!]}{E_{n'{\bf k'}}}\!\Delta_{n'\!{\bf k'}}.
\label{eq:gap-eq}
\end{eqnarray}
Here, $n$ and ${\bf k}$ denote the band index and crystal momentum, respectively, $\Delta$ is the gap function, and $\beta$ is the inverse temperature. The energy $E_{n {\bf k}}$ is defined as $E_{n {\bf k}}$=$\sqrt{\xi_{n {\bf k}}^{2}+\Delta_{n {\bf k}}^{2}}$ and $\xi_{n {\bf k}}=\epsilon_{n {\bf k}}-\mu$ is the one-electron energy measured from the chemical potential $\mu$, where $\epsilon_{n {\bf k}}$ is obtained by solving the normal Kohn-Sham equation in density functional theory (DFT) 
\begin{eqnarray}
\mathcal{H}_{\rm KS}|\varphi_{n{\bf k}}\rangle=\epsilon_{n{\bf k}}
|\varphi_{n{\bf k}}\rangle
\label{eq:KS-eq}
\end{eqnarray}
with $\mathcal{H}_{\rm KS}$ and $|\varphi_{n{\bf k}}\rangle$ being the Kohn-Sham Hamiltonian and the Bloch state, respectively. In Eq.~(\ref{eq:gap-eq}),  $\mathcal{Z}$ and $\mathcal{K}$ are the exchange-correlation kernels in the SCDFT formalism. For the superconducting phase, the self-consistent solution of Eq.(\ref{eq:gap-eq}) becomes nonzero; thereby we determine $T_{\rm c}$. 

\begin{figure}[t!]
\begin{center}
\includegraphics[width=8cm,clip]{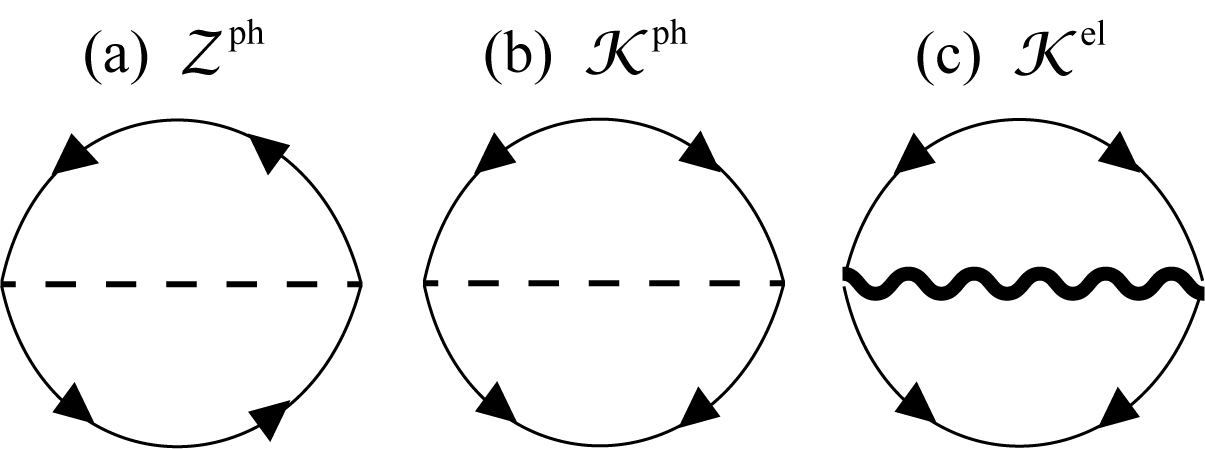}
\end{center}
\caption{Schematic pictures of the exchange-correlation energy to derive the three kernels, $\mathcal{Z}^{\rm ph}$ (a), $\mathcal{K}^{\rm ph}$ (b), and $\mathcal{K}^{\rm el}$ (c). The solid line with arrows in the same (opposite) direction denotes electronic normal (anomalous) propagator. The dashed line denotes phononic propagator, and the bold wavy line denotes the screened electronic Coulomb interaction in the low-frequency limit, respectively.}
\label{fig:diagrams}
\end{figure}

The exchange-correlation kernels $\mathcal{Z}$ and $\mathcal{K}$ are defined by the second-order derivative of the exchange-correlation energy with respect to the electronic anomalous density. In the present study, we consider the three terms~\cite{footnote-diagrams} shown in Fig.~{\ref{fig:diagrams}}, to derive $\mathcal{Z}^{\rm ph}$ [(a)], $\mathcal{K}^{\rm ph}$ [(b)], and $\mathcal{K}^{\rm el}$ [(c)]. Here, $\mathcal{Z}^{\rm ph}$ ($\mathcal{K}^{\rm ph}$) is the lowest-order contribution\cite{Gorling-Levy, Migdal} to the exchange-correlation kernel due to the interaction between phonons and normal (anomalous) electrons, and $\mathcal{K}^{\rm el}$ represents the exchange-correlation interaction between anomalous electrons via screened Coulomb interaction in the low-frequency limit.

These kernels are given in the form averaged over the phonon energy $\omega$ and the electronic energy $\xi'$ ranging from $-\mu$ to $\infty$ as 
\begin{eqnarray}
\mathcal{Z}^{\rm ph}_{n{\bf k}}
&=&
\mathcal{Z}^{\rm ph}(\xi_{n{\bf k}})
\nonumber \\
&=&
- \frac{1}{\mathrm{tanh}[(\beta/2)\xi_{n{\bf k}}]}
\int_{-\mu}^{\infty} d\xi'
    \int d\omega\alpha^2F(\omega)
        \nonumber \\
&&
        \times
        \bigl[
              J(\xi_{n{\bf k}},\xi',\omega)
              +
              J(\xi_{n{\bf k}},-\xi',\omega)
             \bigr]
\label{eq:Z-ph-ene}
\end{eqnarray}
and
\begin{eqnarray}
\mathcal{K}^{\rm ph}_{n{\bf k},n'{\bf k}'}
&=&
\mathcal{K}^{\rm ph}(\xi_{n{\bf k}},\xi'_{n'{\bf k}'})
\nonumber \\
&=&
\frac{2}
     { 
      \mathrm{tanh}[(\beta/2)\xi_{n{\bf k}}]
      \mathrm{tanh}[(\beta/2)\xi'_{n'{\bf k}'}] 
       }
\frac{1}{N(0)}
\nonumber \\
&&
\hspace{0pt}
\times 
\int d\omega
     \alpha^2F(\omega)
     \bigl[ 
      I( \xi_{n{\bf k}},\xi'_{n'{\bf k}'},\omega )
\nonumber \\
&&
\hspace{80pt}
      - I( \xi_{n{\bf k}},-\xi'_{n'{\bf k}'},\omega )
      \bigr],
\label{eq:K-ph-ene}
\end{eqnarray} 
respectively, where $N(0)$ is the electronic density of states (DOS) at the Fermi energy per spin and $\alpha^{2}F(\omega)$ is the Eliashberg function defined as
\begin{eqnarray}
\alpha^{2}F(\omega) = \frac{1}{2\pi N(0)} \sum_{{\bf q}\nu} \frac{\gamma_{{\bf q}\nu}}{\omega_{{\bf q}\nu}} \delta(\omega-\omega_{{\bf q}\nu}).
\label{a2F}
\end{eqnarray}
Here, $\omega_{{\bf q}\nu}$ is the phonon frequency of the mode ${\bf q}$ in the branch $\nu$. Hereafter, we use ${\bf k}$ and ${\bf q}$ to specify the momentum of electrons and phonons, respectively. The $\gamma_{{\bf q}\nu}$ is the phonon linewidth and given as 
\begin{eqnarray}
\gamma_{{\bf q}\nu}=2\pi\omega_{{\bf q}\nu}\sum_{{\bf k}n n'}|g^{{\bf q}\nu}_{{\bf k}+{\bf q}n',{\bf k}n}|^{2}\delta(\xi_{{\bf k}+{\bf q}n'})\delta(\xi_{{\bf k}n}) 
\label{eq:gamma-def}
\end{eqnarray}
with $g^{{\bf q}\nu}_{{\bf k}+{\bf q}n',{\bf k}n}$ being the electron-lattice coupling as 
\begin{eqnarray}
g^{{\bf q}\nu}_{{\bf k}+{\bf q}n',{\bf k}n}
=\langle \varphi_{n'{\bf k+q}} | \delta V_{{\bf q}\nu} | \varphi_{n{\bf k}} \rangle. 
\label{eq:elph-g}
\end{eqnarray}
Here, $\delta V_{{\bf q}\nu}$ is the potential derivative with respect to ${\bf q}\nu$ phonon normal mode, which couples the two electronic states $|\varphi_{n{\bf k}}\rangle$ and $|\varphi_{n'{\bf k+q}}\rangle$. 

The integrand function $I$ in Eq. (\ref{eq:K-ph-ene}) is written as
\begin{eqnarray}
I(\xi,\xi',\omega)
&=&
f_{\beta}(\xi)f_{\beta}(\xi')n_{\beta}(\omega)
\nonumber
\\
&&
\times
\biggl[
 \frac{e^{\beta \xi}-e^{\beta (\xi'+\omega)} }
      {\xi - \xi' - \omega}
 -
 \frac{e^{\beta \xi'}-e^{\beta (\xi+\omega)} }
      {\xi - \xi' + \omega}
 \biggr],
\end{eqnarray}
where $f_{\beta}$ and $n_{\beta}$ are the Fermionic and Bosonic distribution functions, respectively. The integrand function $J$ in Eq. (\ref{eq:Z-ph-ene}) is given as
\begin{eqnarray}
J(\xi,\xi,\omega)
&=&
\tilde{J}(\xi,\xi',\omega)-\tilde{J}(\xi,\xi',-\omega)
\end{eqnarray}
with
\begin{eqnarray}
\tilde{J}(\xi,\xi',\omega)
&=&
     - \frac{
            f_{\beta}(\xi) + n_{\beta}(\omega) 
            }
           {
            \xi - \xi' - \omega
            } 
\biggl[ 
      \frac{
            f_{\beta}(\xi') - f_{\beta}(\xi-\omega) 
            }
           {
            \xi - \xi' - \omega
            }
      \nonumber \\
&&
      -\beta f_{\beta}(\xi-\omega)f_{\beta}(-\xi + \omega)
      \biggr].
\end{eqnarray}

The kernel $\mathcal{K}^{\rm el}$ ascribed to the screened Coulomb interaction between the Cooper pairs is given as 
\begin{eqnarray}
\mathcal{K}^{\rm el}_{n{\bf k},n'{\bf k}'}
&=&
\int d{\bf r}
\int d{\bf r}'
\varphi^{\ast}_{n{\bf k}}({\bf r})
\varphi^{\ast}_{n-{\bf k}}({\bf r}')
W({\bf r},{\bf r}')
\nonumber \\
&&
\times
\varphi_{n'{\bf k}'}({\bf r})
\varphi_{n'-{\bf k}'}({\bf r}'),
\label{eq:Coulomb-def}
\end{eqnarray}
where $W({\bf r},{\bf r}')$ is the static screened Coulomb interaction. In this paper we evaluate this term with two kinds of approximations, i.e., the Thomas-Fermi approximation (TFA) (Ref.~\onlinecite{Marques-metals}) and the random-phase approximation (RPA).~\cite{Sanna-CaC6, Massidda, Hybertsen-Louie} Within TFA, the screened Coulomb interaction is written with the Thomas-Fermi wavenumber $k_{{\rm TFA}}=\sqrt{8\pi N(0)}$ as  
\begin{eqnarray}
W_{\rm TFA}({\bf r},{\bf r}')= 
\frac{e^{-k_{\rm TFA}|{\bf r}-{\bf r}'|}}{|{\bf r}-{\bf r}'|}.
\label{eq:W-TFA}
\end{eqnarray}
 Since $W_{\rm TFA}({\bf r},{\bf r}')$ is given by the sum of its Fourier components as
\begin{eqnarray}
W_{\rm TFA}({\bf r},{\bf r}')
=
\sum_{{\bf k},{\bf G}}\frac{4\pi}{|{\bf k}+{\bf G}|^{2} + k_{\rm TFA}^{2}}e^{{\rm i}({\bf k}+{\bf G})\cdot ({\bf r}-{\bf r}')},
\label{eq:W-TF-Fourier}
\end{eqnarray}
substitution of Eq.~(\ref{eq:W-TFA}) into Eq.(\ref{eq:Coulomb-def}) yields 
\begin{eqnarray}
\mathcal{K}^{\rm el,TFA}_{n{\bf k},n'{\bf k}'}
&=&
\frac{4\pi}{\Omega}
\sum_{\bf G}
\frac{1}{|{\bf k}\hspace{-3pt}-\hspace{-3pt}{\bf k}'\hspace{-3pt}+\hspace{-3pt}{\bf G}|^{2}\hspace{-2pt}+\hspace{-2pt}k^{2}_{\rm TFA}}
|\rho^{n'{\bf k}'}_{n{\bf k}}({\bf G})|^{2}
\label{eq:K-el-TF}
\end{eqnarray}
with
\begin{eqnarray}
\rho^{n'{\bf k}'}_{n{\bf k}}({\bf G})
&=&
\int_{\Omega} d{\bf r}
\varphi^{\ast}_{n'{\bf k}'}({\bf r})
e^{{\rm i}({\bf k}'-{\bf k}+{\bf G})\cdot{\bf r}}
\varphi_{n{\bf k}}({\bf r}), 
\label{eq:rho}
\end{eqnarray}
where ${\bf G}$ is the reciprocal lattice vector, $\Omega$ is the volume of the unit cell, and $\int_{\Omega}$ denotes the integration within the unit cell. In this approximation, the screened interaction is isotropic and the screening length is determined by the electronic DOS at the Fermi energy. This interaction is nothing but that derived from the RPA in the vacuum and in the long-wavelength limit, where the local field effect of crystal is totally ignored in the calculation of screening.

We next consider the RPA expression of the screened Coulomb interaction in the low-frequency limit. In general, the screened Coulomb interaction of crystal is expressed with the double-Fourier transform as\cite{Hybertsen-Louie}
\begin{eqnarray}
\hspace{-8pt}
W({\bf r},{\bf r}')
&=& 
\sum_{{\bf K}{\bf G}{\bf G}'}
e^{{\rm i}({\bf K}+{\bf G})\cdot {\bf r}}
W_{{\bf G}{\bf G}'}({\bf K})
e^{-{\rm i}({\bf K}+{\bf G}')\cdot {\bf r}'} 
\label{eq:W-RPA-Fourier}
\end{eqnarray}
with
\begin{eqnarray}
W_{{\bf G}{\bf G}'}({\bf K})
&=&
4\pi
\frac{1}{|{\bf K}+{\bf G}|}
\tilde{\varepsilon}^{-1}_{{\bf G}{\bf G}'}({\bf K})
\frac{1}{|{\bf K}+{\bf G}'|}.
\label{WRPAGG}
\end{eqnarray}

Here, ${\bf K}$ is the wavevector in the first Brillouin zone and $\tilde{\varepsilon}_{{\bf G}{\bf G}'}({\bf K})$ is the symmetrized dielectric matrix,\cite{Hybertsen-Louie} which is calculated using the following expression
\begin{eqnarray}
\tilde{\varepsilon}_{{\bf G}{\bf G}'}({\bf K})
&=&
\delta_{{\bf G}{\bf G}'}-4\pi\frac{1}{|{\bf K}+{\bf G}|}\chi^{0}_{{\bf G}{\bf G}'}({\bf K})\frac{1}{|{\bf K}+{\bf G}'|}
\end{eqnarray}
with $\chi^{0}$ being the independent-particle polarization as 
\begin{eqnarray}
\chi^{0}_{{\bf G}{\bf G}'}({\bf K})
&=&
\frac{2}{\Omega}
\sum_{n,n',{\bf k}}
 \frac{f_{\beta}(\xi_{n{\bf k}})\bigl\{1-f_{\beta}(\xi_{n'{\bf k}+{\bf K}})\bigr\}}{\xi_{n{\bf k}} - \xi_{n'{\bf k}+{\bf K}}}
 \nonumber \\
&&
\hspace{-5pt}
 \times
\bigl[\{\rho^{n'{\bf k}+{\bf K}}_{n{\bf k}}({\bf G})\}^{\ast}
\rho^{n'{\bf k}+{\bf K}}_{n{\bf k}}({\bf G}')
+ {\rm c. c.}\bigr].
\label{eq:chi-def}
\end{eqnarray}
After inserting Eq.~(\ref{eq:W-RPA-Fourier}) into Eq.~(\ref{eq:Coulomb-def}) and with ${\bf K}$=${\bf k}$$-$${\bf k'}$, we obtain 
\begin{eqnarray}
\mathcal{K}^{\rm el,RPA}_{n{\bf k},n'{\bf k}'}\!=\!
\frac{4\pi}{\Omega}\!
\sum_{{\bf G}\!{\bf G}'}\! 
\frac{
\!\rho^{n{\bf k}}_{n'{\bf k}'}(\!{\bf G}\!)\tilde{\varepsilon}^{-1}_{{\bf G}{\bf G}'}(\!{\bf k}\!-\!{\bf k}'\!)\!\{\rho^{n{\bf k}}_{n'{\bf k}'}(\!{\bf G}'\!)\!\}^*
}
{
|{\bf k}-{\bf k}'+{\bf G}||{\bf k}-{\bf k}'+{\bf G}'|
}\!.
\label{eq:K-el-RPA}
\end{eqnarray}
The resulting kernel $\mathcal{K}^{\rm el,RPA}$ includes the local-field effect on the screening in the real crystal. In this expression, $\mathcal{K}^{\rm el,RPA}$ has a singularity at ${\bf k}$=${\bf k}'$ and ${\bf G}$=${\bf 0}$ or ${\bf G}'$=${\bf 0}$. This singularity is avoided following Ref.~\onlinecite{Pick-ASR}.

\begin{figure}[h!]
\begin{center}
\includegraphics[scale=0.40]{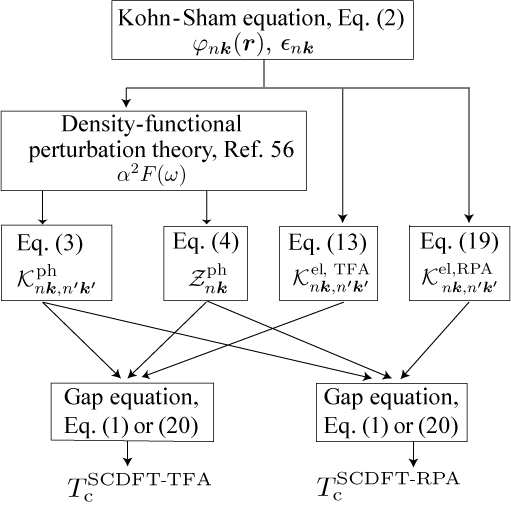}
\caption{Procedure to obtain the superconducting transition temperatures through density functional theory for superconductors.}
\label{fig:SCDFT-flowchart}
\end{center}
\end{figure}

Figure \ref{fig:SCDFT-flowchart} describes the flowchart of the present SCDFT calculations. Starting from solving the Kohn-Sham equation in Eq.~(\ref{eq:KS-eq}), we next perform density-functional perturbation calculations to obtain $\alpha^{2} F(\omega)$, which generates electron-phonon kernels $\mathcal{Z}^{{\rm ph}}$ [Eq.~(\ref{eq:Z-ph-ene})] and $\mathcal{K}^{{\rm ph}}$ [Eq.~(\ref{eq:K-ph-ene})]. Also, we calculate the electron-electron kernel $\mathcal{K}^{{\rm el}}$ for TFA with Eq.~(\ref{eq:K-el-TF}) and for RPA with Eq.~(\ref{eq:K-el-RPA}). With these information as inputs, the gap equation in Eq.~(\ref{eq:gap-eq}) is solved. According to the two types of $\mathcal{K}^{{\rm el, TFA}}$ and $\mathcal{K}^{{\rm el, RPA}}$, two values of $T_{\rm c}$ are estimated, which are denoted as $T_{\rm c}^{\rm SCDFT\mathchar`-TFA}$ and $T_{\rm c}^{\rm SCDFT\mathchar`-RPA}$. From comparison between $T_{\rm c}^{\rm SCDFT\mathchar`-TFA}$ and $T_{\rm c}^{\rm SCDFT\mathchar`-RPA}$, we study the effects beyond the uniform and local electronic screening on the transition temperature.

We remark some technical details when we solve the gap equation in Eq.(\ref{eq:gap-eq}). In the present calculation, the states in the gap equation, labeled by $n{\bf k}$, are generated by sampling; they are generated so that their density distributes logarithmically in the energy range of interest. For this purpose, with a given band $n$, the Brillouin zone (BZ) is divided into four regions, following a criterion for ${\bf k}$; (i) $|\xi_{n{\bf k}}| < \omega_{0}$, (ii) $\omega_{0}< |\xi_{n{\bf k}}| < \omega_{1}$, (iii) $\omega_{1} < |\xi_{n{\bf k}}| < \omega_{2}$, (iv) $ \omega_{2} < |\xi_{n{\bf k}}|$, where $\xi_{n{\bf k}}$ is the energy of the sampling state and $\omega_{i}$ ($i$$=$$0,1,2$) are the energy criteria. In the present case, $\omega_{0}$, $\omega_{1}$, and $\omega_{2}$ are set to $0.002$ eV, $0.02$ eV, and $0.35$ eV, respectively. Next, to realize logarithmic distribution of the sampled states, we introduce an acceptance ratio $p$ for each region; $p$=1 for (i), $p$=0.1 for (ii), $p$=0.01 for (iii) and $p$=0.002 for (iv). With this acceptance ratio, the randomly-generated sampling states form a logarithmic distribution. For the resulting $n{\bf k}$ points, the energy $\xi_{n{\bf k}}$ and the exchange-correlation kernels $\mathcal{Z}_{n{\bf k}}$ and $\mathcal{K}_{n{\bf k},n'{\bf k}'}$ are evaluated with the linear tetrahedron interpolation scheme\cite{tetrahedra,comment-kernel} for the original {\em ab initio} data. On the ground that the density of the sampling points is not uniform, the gap equation in Eq.(1) is rewritten as
\begin{eqnarray}
\hspace{-20pt}
\Delta_{n{\bf k}}
&=&
-\mathcal{Z}_{n{\bf k}}\Delta_{n{\bf k}}
\nonumber \\
&&
\hspace{-2pt}
-\frac{1}{2}
\sum_{n'{\bf k'}}\hspace{-3pt}W_{n'{\bf k}'}
 \mathcal{K}_{n{\bf k},n'{\bf k}}
\frac{ 
  \mathrm{tanh}[
       (\beta/2)E_{n'{\bf k'}}
        ]
       }
     { 
      E_{n'{\bf k'}}
       }
\Delta_{n'{\bf k'}}
\label{eq:gap-eq-aux}
\end{eqnarray}
with $W_{n{\bf k}}$$=$$V_{n\alpha}/N_{n\alpha}$ being the weight normalized as $\sum_{\bf k}W_{n{\bf k}}$$=$$1$ for each $n$. Here, $\alpha$ specifies the region where the sampling state $n{\bf k}$ belongs, and $V_{n\alpha}$ is the volume of the region $\alpha$ in the first BZ for band $n$, estimated by the frequency of sampling.\cite{Kawamura-priv} The total number of the accepted sampling points, $N_{n\alpha}$, also depends on $\alpha$ and $n$. We found that 10,000 sampling states per band crossing the Fermi energy assure a convergence within a few percent.

\begin{figure}[b!]
\begin{center}
\includegraphics[scale=0.41]{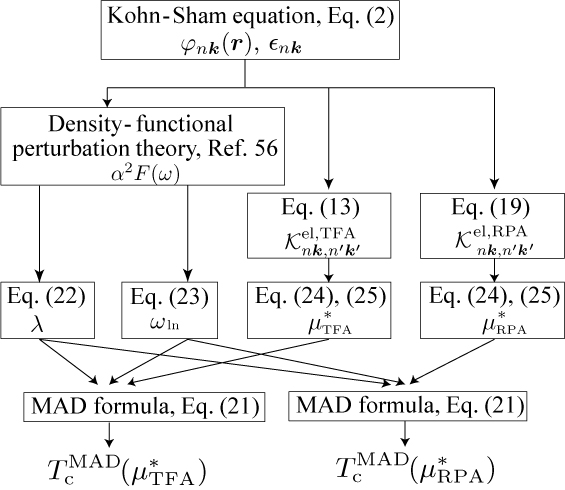}
\caption{Procedure to obtain the superconducting transition temperatures based on the McMillan-Allen-Dynes formula.} 
\label{fig:MAD-flowchart}
\end{center}
\end{figure}
As a reference, we also calculate $T_{\rm c}$ using the McMillan-Allen-Dynes formula (MAD)
\begin{eqnarray}
T_{\rm c}=
\frac{\omega_{\rm ln}}{1.2}
{\rm exp}\biggl[-\frac{1.04(1+\lambda)}
          {\lambda - \mu^{\ast}(1+0.62\lambda)}
    \biggr].
\label{eq:McMillan}
\end{eqnarray}
The parameters $\lambda$ and $\omega_{\rm ln}$ are defined using the Eliashberg function $\alpha^{2}F$ as 
\begin{eqnarray}
\lambda=2\int d\omega \frac{\alpha^{2}F(\omega)}{\omega}
\label{eq:lambda-def}
\end{eqnarray}
and
\begin{eqnarray}
\omega_{\rm ln}=
{\rm exp}\Biggl[
          \frac{
          \int d\omega \frac{\alpha^{2}F(\omega)}{\omega} {\rm ln}\omega
               }
               {
          \int d\omega \frac{\alpha^{2}F(\omega)}{\omega}               
               }
          \Biggr],  
\label{eq:omegaln-def}
\end{eqnarray}
respectively. The parameter $\mu^{\ast}$ in Eq.~(\ref{eq:McMillan}) is an effective Coulomb pseudopotential in the narrow energy region within the Debye frequency $\omega_{{\rm D}}$ above/below the Fermi level. Within RPA,~\cite{Lee-Chang-Cohen}  $\mu^{\ast}$ is written with the renormalization formula\cite{Morel-Anderson} as
\begin{eqnarray}
\mu^{\ast} = 
\frac{\mu}
     {1+\mu{\rm ln}\frac{E_{\rm el}}{\omega_{\rm D}}}, 
\label{eq:renorm}
\end{eqnarray}
where $E_{\rm el}$ is a parameter defining the electronic energy scale. It is conventionally set to the Fermi energy measured from the conduction-band bottom, assuming that the band dispersion is parabolic. The bare $\mu$ is defined as 
\begin{eqnarray}
\mu=N(0) \langle\!\langle \mathcal{K}^{\rm el} \rangle\!\rangle_{\rm FS},
\label{eq:mu-def}
\end{eqnarray}
where $\langle\!\langle \mathcal{K}^{\rm el} \rangle\!\rangle_{\rm FS}$ is the average of the electronic interaction over the Fermi surface and is defined as 
\begin{eqnarray} 
\langle\!\langle \mathcal{K}^{\rm el} 
\rangle\!\rangle_{\rm FS}
=\frac{1}{N(0)^{2}}\sum_{n{\bf k},n'{\bf k}'} 
 \delta(\xi_{n{\bf k}}) \delta(\xi_{n'{\bf k}'}) 
 \mathcal{K}^{\rm el}_{n{\bf k},n'{\bf k}'}. 
\label{KelFS} 
\end{eqnarray}
 
Since $\mu$ is estimated for both of $\mathcal{K}^{{\rm el, TFA}}$ [Eq.~(\ref{eq:K-el-TF})] and $\mathcal{K}^{{\rm el, RPA}}$ [Eq.~(\ref{eq:K-el-RPA})], we distinguish them as $\mu_{{\rm TFA}}$ and $\mu_{{\rm RPA}}$. Thus, we derive the two $T_{\rm c}$ values, i.e., $T_{\rm c}^{\rm MAD}(\mu^{\ast}_{\rm TFA})$ and $T_{\rm c}^{\rm MAD}(\mu^{\ast}_{\rm RPA})$. We summarize in Fig.\ref{fig:MAD-flowchart} the procedure of the MAD calculations. 

In the MAD analysis, the quantitative reliability of the resulting $T_c$ is somewhat ambiguous, because the renormalization formula in Eq.~(\ref{eq:renorm}) includes an empirical parameter $E_{{\rm el}}$. An advantage of the SCDFT framework over the MAD analysis is that the former does not contain any empirical parameters such as $E_{{\rm el}}$. In the present paper, we study quantitatively the electronic-interaction effect on $T_c$ with systematic comparisons between SCDFT and MAD, and between TFA and RPA.

\section{Simple Metals}\label{sec:metal}
To check the reliability of our SCDFT code, we applied it to benchmark systems; aluminum and niobium, which are weak-coupling and strong-coupling BCS superconductors, respectively. The comparison between our results and experiments, together with other SCDFT result based on TFA,~\cite{Marques-metals} are shown in Table \ref{tab:elemental-metal}. The table also contains the $T_{c}$ values based on the MAD formula and several key parameters used in the MAD calculations.
\begin{table}[b!]
\caption{Our calculated superconducting transition temperature $T_c$ based on density functional theory for superconductors (SCDFT), together with the result based on the McMillan-Allen-Dynes (MAD) formula in Eq.(\ref{eq:McMillan}). Experimental and previous SCDFT results are listed for comparison. Parameters used in the MAD calculations are also given. The $\mu_{{\rm RPA}}$ and $\mu_{{\rm RPA}}^{*}$ parameters are compared with the other theoretical results.}
\begin{center}
\label{tab:elemental-metal}
{\tabcolsep = 3mm
\begin{tabular}{lrr} \toprule[2pt]
 & Al & Nb \\
\midrule[0.5pt]
$T_{\rm c}^{\rm SCDFT\mathchar`-TFA}$ [K]  & 0.79  &  9.9  \\
$T_{\rm c}^{\rm SCDFT\mathchar`-RPA}$ [K]  & 0.72  &  8.5  \\
$T_{\rm c}^{\rm MAD}(\mu^{\ast}_{\rm TFA})$ [K]  & 1.49  & 15.1  \\
$T_{\rm c}^{\rm MAD}(\mu^{\ast}_{\rm RPA})$ [K]  & 1.43  & 14.5  \\
\midrule[0.2pt]
$T_{\rm c,ref}^{\rm SCDFT\mathchar`-TFA}$ [K]  &  0.90\footnotemark[1]  &  9.5\footnotemark[1]  \\
$T_{\rm c}^{\rm expt}$ [K]  & 1.20\footnotemark[2]  & \ph 9.26\footnotemark[2] \\
\midrule[0.5pt]
$\lambda$                        & 0.419  & 1.305  \\
$\omega_{\rm ln}$ [K]            & 308   & 164   \\
$\mu_{\rm TFA}$            &0.257  &0.329  \\
$\mu_{\rm RPA}$           & 0.269 & 0.433 \\
$\mu^{\ast}_{\rm TFA}$            & 0.105 & 0.119 \\
$\mu^{\ast}_{\rm RPA}$           & 0.107 & 0.130 \\
\midrule[0.2pt]
$\mu_{\rm RPA, ref}$       & 0.236\footnotemark[3] & 0.488\footnotemark[4] \\
$\mu^{\ast}_{\rm RPA, ref}$& 0.100\footnotemark[3] & 0.133\footnotemark[4] \\
\midrule[2pt]
\multicolumn{3}{l}{{\scriptsize $^{a}$SCDFT result with $\mathcal{K}^{\rm el,TFA}$ in Eq.~(\ref{eq:K-el-TF}) taken}}\\[-0.4mm]
\multicolumn{3}{l}{{\scriptsize from Ref.\onlinecite{Marques-metals}.}}\\[-0.4mm]
\multicolumn{3}{l}{{\scriptsize $^{b}$Experimental values taken from Ref.\onlinecite{Ashcroft}.}}\\[-0.4mm]
\multicolumn{3}{l}{{\scriptsize $^{c}${\em Ab initio} RPA result in Ref.\onlinecite{Lee-Chang-Cohen}.}}\\[-0.4mm]
\multicolumn{3}{l}{{\scriptsize $^{d}${\em Ab initio} RPA result in Ref.\onlinecite{Lee-Chang-Nb}.}}\\[-0.4mm]
\end{tabular}
}
\end{center}
\end{table}

Our calculated $T_{\rm c}^{\rm SCDFT\mathchar`-TFA}$ and $T_{\rm c}^{\rm SCDFT\mathchar`-RPA}$ show a good agreement with the previous calculation ($T_{\rm c,ref}^{\rm SCDFT\mathchar`-TFA}$) and the experiments ($T_{\rm c}^{\rm expt}$). The small deviation between the present and previous calculations comes from minor differences in $\alpha^{2}F$ used for the calculation of the phononic kernels. It is interesting to note that the MAD formula significantly overestimates $T_{\rm c}$ [see $T_{\rm c}^{\rm MAD}(\mu^{\ast}_{\rm TFA})$ and $T_{\rm c}^{\rm MAD}(\mu^{\ast}_{\rm RPA})$ in the table]. This error can partly be attributed to the oversimplification of the electronic-interaction kernel in MAD, i.e., the use of $\mu$ in Eq.~(\ref{eq:mu-def}),~\cite{comment-approx} ignoring the wavenumber dependence of the screened Coulomb interaction. In this approximation, the Coulomb interaction is approximated by a $\delta$ function in the real space, so that electrons forming the Cooper pair do not feel distant Coulomb repulsion. As a result, the electronic interaction is estimated to be smaller in MAD than SCDFT considering the wavenumber dependence of the Coulomb interaction and $T_{c}$ obtained by the MAD formula tends to be higher than $T_{c}$ estimated within SCDFT.  We also note that the difference between TFA and RPA is small, which indicates that the local-field effect is not so significant for Al and Nb.

We note that the deviation of our $T^{\rm SCDFT}_{\rm c}$ from the experiment for Al is relatively large if we consider the $T_c$ ratio. As suggested in Ref.\onlinecite{Marques-metals}, in the superconductors with small gap with $T_{\rm c}$ below a few K, the $T_{\rm c}$ value is a consequence of a subtle balance between the electron-phonon coupling and the screened electron-electron interaction, thereby systematic errors depending on the computational details of approximations are strongly enhanced. This type of error arising from a subtraction of two nearly equal quantities does not occur in the superconductors with relatively high $T_{\rm c}$ and do not need to be worried in the following sections.

\section{Layered nitride chlorides}\label{sec:result}
\subsection{Computational Detail}
{\em Ab initio} electronic and lattice-dynamical calculations were performed for lithium-doped layered metallonitride chrolides, $\beta$-Li$_{x}$MNCl (M=Ti, Zr, and Hf) with the three doping rate $x$=0.0, 0.3, and 0.5 with {\sc Quantum Espresso} package,\cite{QE} where the local density approximation (LDA) with the parameterization by Perdew and Zunger~\cite{PZ81, Ceperley-Alder} and the Troullier-Martins norm-conserving pseudopotentials\cite{Troullier-Martins} were employed. The pseudopotentials for Ti, Zr, and Hf were generated in the semicore configurations of (3$s$)$^{2.0}$(3$p$)$^{6.0}$(3$d$)$^{2.0}$, (4$s$)$^{2.0}$(4$p$)$^{6.0}$(4$d$)$^{2.0}$, and (5$s$)$^{2.0}$(5$p$)$^{6.0}$(5$d$)$^{2.0}$, respectively. The scalar-relativistic correction\cite{Koelling-Harmons} was applied to the Zr and Hf pseudopotentials. The Li pseudopotentials were supplemented with the partial core correction.\cite{Louie-pcc} The lattice parameters and internal coordinates were fully optimized. We checked that our calculations for the relaxed structures and the $\Gamma$-point phonon frequencies well reproduce the experiments (Appendix~\ref{app:Gammaphon}). 

The cutoff energies in the wavefunctions were set to 75 Ry, 80 Ry, and 90 Ry in the Ti, Zr, and Hf compounds, respectively. The charge density was described with the 8$\times$8$\times$8 $k$ points in the Monkhorst-Pack grids\cite{MP-grid} and the electronic DOS was calculated using the denser 32$\times$32$\times$32 $k$ grids. Phonon dynamical matrices were calculated with the 4$\times$4$\times$4 $q$ points, while the electron-phonon couplings \{$g^{{\bf q}\nu}_{{\bf k}+{\bf q}n',{\bf k}n}$\} in Eq. (\ref{eq:elph-g}) were calculated with the 4$\times$4$\times$4 $q$ phonon modes and the 32$\times$32$\times$32 $k$ wavefunctions. The polarization matrix \{$\chi^{0}_{{\bf G}{\bf G}'}({\bf K})$\} in Eq.~(\ref{eq:chi-def}) was expanded with the plane waves with the cutoff of 12.8 Ry and calculated on the 5$\times$5$\times$5 $k$ grid. The Kohn-Sham states within a range [$-$12 eV, 36 eV] were included in the polarization calculation. The Fermi-surface integrals involving $\gamma_{{\bf q}\nu}$ in Eq.~(\ref{eq:gamma-def}) and $\mu$ in Eq.~(\ref{eq:mu-def}) were performed using the first-order and zeroth-order of the Hermite-Gaussian smearing,\cite{Methfessel-Paxton} respectively. The smearing width was set to 0.020 Ry in $x$=0.5 and 0.025 Ry in $x$=0.3. The SCDFT calculation in Eq.~(\ref{eq:gap-eq-aux}) was done with 10,000 sampling $k$ points for bands crossing the Fermi energy and 500 points for the others, where we considered 25 valence and 25 conduction bands. 

\subsection{Atomic geometry}
\begin{figure}[b!]
\begin{center}
\includegraphics[scale=0.23]{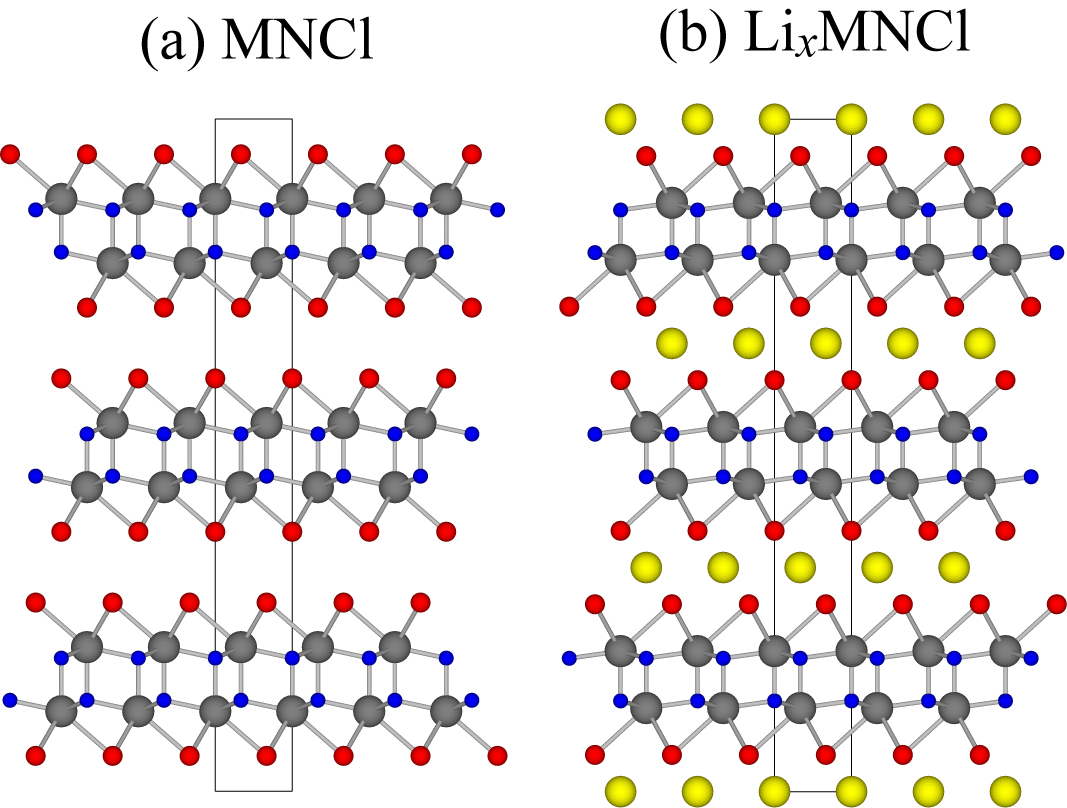}
\caption{(Color online) Atomic geometry of $\beta$-ZrNCl (a) and $\beta$-Li$_{x}$ZrNCl (b) along the $a$ axis, displayed with the conventional cell (solid line). Gray, blue, red, and yellow spheres stand for Zr, N, Cl, and Li atoms, respectively. The space group is {\it R}3{\it m} (No. 166). The Li atom is located at the 3{\it a} site in terms of the Wyckoff position, whereas the other atoms are located at the 6{\it c} sites, which are specified by one parameter $z$ (see also Table. \ref{tab:str-nondoped} and \ref{tab:str-doped}).}
\label{fig:struct}
\end{center}
\end{figure}
We show in Fig.~\ref{fig:struct} atomic structures of $\beta$-ZrNCl (a) and $\beta$-Li$_{0.5}$ZrNCl (b). We display the side view along the $a$ axis. The atomic structure consists of stacking block layers made from Zr (gray), N (blue), and Cl (red) atoms. On inserting Li (yellow) in between the block layers, the crystal slightly expands. Here the positions of the Zr, N, and Cl atoms, as well as the intercalated Li atom have been determined with experiments.\cite{Shamoto-stacking, Fuertes, Istomin, Adelmann, Chen-nondoped, Chen-doped} It is also experimentally confirmed that the stacking order of the layers for doped systems is different from that of the mother compounds.\cite{Shamoto-stacking} The Hf compounds basically take the same structure.\cite{Fuertes,Shamoto2004} The details of the atomic configuration are found in Ref.~\onlinecite{Yamanaka-review}.

We summarize in Table \ref{tab:str-nondoped} the optimized structural parameters for the non-doped system $\beta$-MNCl. The lattice parameters are approximately 3\% smaller than the experiment and the internal $z$ parameters agree well with it. Although the Ti compound with the $\beta$ structure is not synthesized experimentally, in the present paper, we study this material as a reference and compare it with other materials. The optimized internal parameters of the Ti compound are very similar to those of the Zr and Hf compounds, whereas the lattice parameters are somewhat smaller, which is reasonably understood in terms of smallness of the ionic radius of Ti.

\begin{table}[t!]
\caption[t]{Our calculated structural parameters for undoped $\beta$-MNCl.}
\begin{center}
\label{tab:str-nondoped}
{\scriptsize 
\begin{tabular}{lc c ccc c cc} \toprule[2pt] 
M& Ti & & \multicolumn{3}{c}{Zr} & & \multicolumn{2}{c}{Hf} \\
\cline{2-2}\cline{4-6}\cline{8-9} 
  & present & 
  & present & expt.$^{a}$ & 
  theory$^{b}$ & 
  & present & expt.$^{a}$ \\
\midrule[0.5pt]
$a$[\AA] &\ph 3.342   & &\ph 3.553  &\ph 3.605  &\ph 3.556  & &\ph 3.488  &\ph 3.577  \\
$c$[\AA] &   26.164   & &   26.879  &   27.672  &   27.178  & &   26.762  &   27.711  \\
$z$(M)   &\ph 0.120   & &\ph 0.119  &\ph 0.119  &\ph 0.118  & &\ph 0.119  &\ph 0.120 \\
$z$(N)   &\ph 0.197   & &\ph 0.198  &\ph 0.198  &\ph 0.198  & &\ph 0.198  &\ph 0.198  \\
$z$(Cl)  &\ph 0.388   & &\ph 0.386  &\ph 0.388  &\ph 0.386  & &\ph 0.387  &\ph 0.388 \\
\midrule[2pt]
\multicolumn{9}{l}{$^{a}$Single-crystal X-ray diffraction measurement}\\
\multicolumn{9}{l}{taken from Ref.~\onlinecite{Chen-nondoped}.}\\
\multicolumn{9}{l}{$^{b}${\it Ab initio} DFT-LDA full-potential result in Ref.~\onlinecite{Sugimoto-Oguchi}.}\\
\end{tabular} 
}
\end{center}
\end{table}

We next show in Table~\ref{tab:str-doped} our calculated structural parameters for the doped system $\beta$-Li$_{x}$MNCl.  For the Zr compounds, the theoretical parameters agree well with experiments. For the Hf compounds, the experimental $c$ parameter is somewhat larger than the theoretical value, which may originate from the difference in the intercalated atoms, Li and Na. For the reliable description of the atomic structure of the incommensurate filling $x$=0.3, the virtual crystal approximation that the Li ionic charge is artificially reduced from the nominal value of 3 to 2.6 was employed. This is critical to reproduce the experiment; a conventional treatment of electron doping with a uniform compensating positive background charge leads to an inaccurate description for the lattice parameters, a deviation from the experiment by $\sim$10 \%. In our virtual crystal approximation, the error is $\sim$1\% for the Zr compound.
\begin{table}
\caption[t!]{Our calculated structural parameters for doped $\beta$-Li$_{x}$MNCl.}
\begin{center}
{\scriptsize 
\label{tab:str-doped}
\begin{tabular}{lcccccccccc} \toprule[2pt]
M & \multicolumn{2}{c}{Ti} & & \multicolumn{3}{c}{Zr} & & \multicolumn{3}{c}{Hf} \\
\cline{2-3}\cline{5-7}\cline{9-11} 
 & \multicolumn{2}{c}{present} & & \multicolumn{2}{c}{present} & expt.$^{a}$ & & \multicolumn{2}{c}{present} & expt.$^{b}$ \\
\cline{2-3}\cline{5-6}\cline{7-7}\cline{9-10}\cline{11-11} 
$x$ & 0.3 & 0.5 & & 0.3 & 0.5 & 0.20 & & 0.3 & 0.5 &0.29 \\
\midrule[0.5pt]
$a$[\AA] &\ph 3.394 &\ph 3.406& &\ph 3.604   &\ph 3.616   &\ph 3.591  & &\ph 3.541   &\ph 3.561  &\ph 3.589  \\
$c$[\AA] &   27.740 &   27.524& &   28.163   &   27.942   &   27.839  & &   28.173   &   27.795  &   29.722 \\
$z$(M)   &\ph 0.207 &\ph 0.206& &\ph 0.209   &\ph 0.208   &\ph 0.213  & &\ph 0.208   &\ph 0.207  &\ph 0.208  \\
$z$(N)   &\ph 0.136 &\ph 0.134& &\ph 0.135   &\ph 0.132   &\ph 0.136  & &\ph 0.135   &\ph 0.133  &\ph 0.137  \\
$z$(Cl)  &\ph 0.391 &\ph 0.385& &\ph 0.388   &\ph 0.383   &\ph 0.389  & &\ph 0.389   &\ph 0.384  &\ph 0.394  \\
\midrule[2pt]
\multicolumn{11}{l}{$^{a}$Single-crystal X-ray diffraction measurement in Ref.~\onlinecite{Chen-doped}.}\\
\multicolumn{11}{l}{$^{b}$Powder neutron diffraction measurement for $\beta$-Na$_{x}$HfNCl in Ref.~\onlinecite{Shamoto2004}.}\\
\end{tabular}
}
\end{center}
\end{table}

\subsection{Electronic Bandstructure}
\begin{figure*}[t!]
\begin{center}
\includegraphics[scale=0.37]{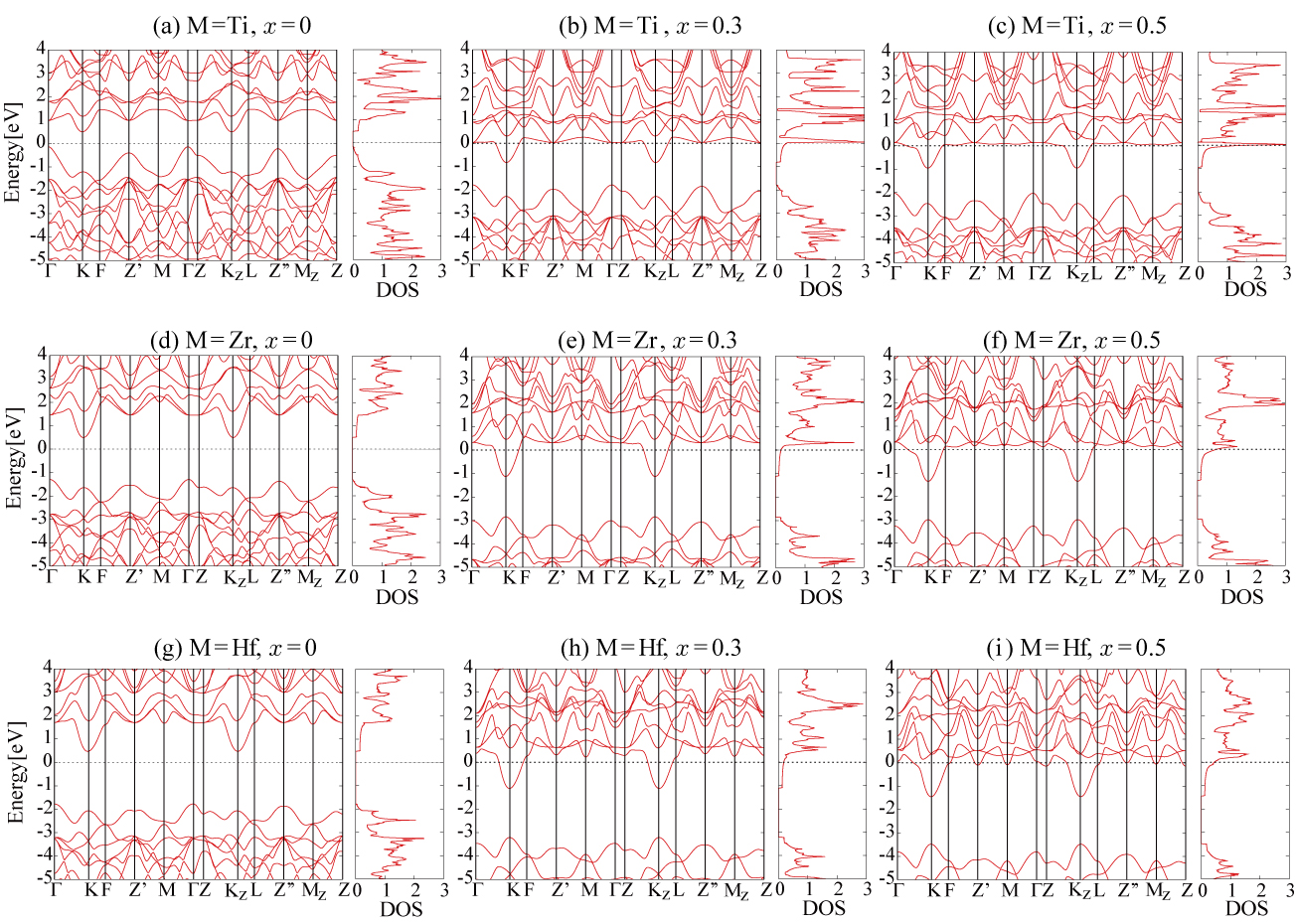}
\caption{(Color online) Our calculated band structures and density of states (DOS) of $\beta$-Li$_{x}$MNCl ($x$=0, 0.3, and 0.5; M=Ti, Zr, and Hf). The DOS is given in the unit of eV$^{-1}$spin$^{-1}$(f.u.)$^{-1}$. The DOS was calculated with the tetrahedron integration method\cite{tetrahedra} for the electronic bands interpolated onto 32$\times$32$\times$32 $k$ mesh. Dashed lines represent the Fermi energy. We employed a conventional set of special points following Ref.~\onlinecite{Heid-DFPT}. The points K$_{{\rm Z}}$ and M$_{{\rm Z}}$ stand for the K+Z and M+Z points, respectively.}
\label{fig:bands-dos}
\end{center}
\end{figure*}

We show in Fig.~\ref{fig:bands-dos} our calculated band structure and electronic DOS of $\beta$-Li$_x$MNCl ($x$=0, 0.3, and 0.5; M=Ti, Zr, and Hf) for the range of [$-$5 eV, 4 eV]. The undoped system ($x$=0) is an insulator with a band gap of $\sim$0.5 eV for M=Ti and $\sim$2 eV for M=Zr and Hf. Upon doping (from the left to right panels), we see that carriers are accommodated in the conduction bottom, without changing its two-dimensional dispersion significantly. While the DOS of this two-dimensional band is almost constant as a function of energy, $N(0)$ increases (see Table \ref{tab:metaltable}) by carrier doping. For the material dependence (from the top to bottom panels), we see that the band structures have several common features; for example, the conduction bottom always locates at the K point. On the other hand, DOS around the Fermi level is larger for the lighter element systems. In particular, the Ti compounds have high $N(0)$ because the van-Hove singularity is closer to the conduction bottom, which seems favorable to have a strong electron-phonon coupling.
\begin{table}[b!]
\caption[t]
{Density of states $N(0)$ at the Fermi energy [/(eV spin f.u.)] and parameters in the McMillan-Allen-Dynes formula in Eq.(\ref{eq:McMillan}). For $N(0)$, $\lambda$, and $\omega_{\rm ln}$, our calculated results are compared with the experimental and previous theoretical results.} 
\begin{center}
\label{tab:metaltable}
\begin{tabular}{lccccccccc} \toprule[2pt]
M & & \multicolumn{2}{c}{Ti} & & \multicolumn{2}{c}{Zr} & &\multicolumn{2}{c}{Hf} \\
 \cline{3-4}\cline{6-7}\cline{9-10}
$x$ & & 0.3 & 0.5 & & 0.3 & 0.5 & & 0.3 & 0.5 \\
\midrule[0.5pt]
$N(0) $ &  &0.342 &1.675 & &0.191 &0.662 & &0.176 &0.437 \\
$N_{\rm expt.}(0) $ &  & -- & -- & &\multicolumn{2}{c}{0.23$^{a}$} & &\multicolumn{2}{c}{0.25$^{b}$} \\
$N_{\rm LDA}(0) $ & & -- & -- &  &\multicolumn{2}{c}{0.169--0.146$^{c}$} & &\multicolumn{2}{c}{0.19$^{d}$} \\
\midrule[0.2pt]
$\lambda$                                     &  &0.286  &0.411 & &0.552  &0.982 & &0.827  &1.292 \\
$\lambda_{\rm expt.}$                         &  & -- & -- &  &\multicolumn{2}{c}{0.22$^{a}$} & & -- & --  \\
$\lambda_{\rm LDA}$                           &  & -- & -- & &\multicolumn{2}{c}{0.521--0.508$^{c}$} & & -- & -- \\
\midrule[0.2pt]
$\omega_{\rm ln}$ [K]                         &  &459  &429 &  &421  &305 & &324 &204  \\
$\omega_{\rm LDA}$[K]                        &  & -- & -- & &\multicolumn{2}{c}{422--415$^{c}$} & & -- & -- \\
\midrule[0.2pt]
$\mu_{\rm TFA}$                                &  &0.771 &1.705 & &0.329 &0.576 & &0.413 &0.367 \\
$\mu_{\rm RPA}$                               &  &0.238 &1.066 & &0.161 &0.441 & &0.161 &0.263 \\
$\mu^{\ast}_{\rm TFA}$                         &  &0.281 &0.331 & &0.177 &0.218 & &0.200 &0.178 \\
$\mu^{\ast}_{\rm RPA}$                        &  &0.155 &0.296 & &0.113 &0.196 & &0.114 &0.149 \\
\midrule[2pt]
\multicolumn{10}{l}{$^{a}${\scriptsize Specific-heat measurement taken from Ref.\onlinecite{Taguchi-specheat}.}}\\[-0.4mm]
\multicolumn{10}{l}{$^{b}${\scriptsize Magnetic susceptibility measurement taken from Ref.\onlinecite{Tou-suscep}.}}\\[-0.4mm]
\multicolumn{10}{l}{$^{c}${\scriptsize {\it Ab initio} DFT-LDA result for $\beta$-Li$_{1/6}$ZrNCl in Ref.\onlinecite{Heid-DFPT}.}}\\[-0.4mm]
\multicolumn{10}{l}{$^{d}${\scriptsize {\it Ab initio} DFT-LDA result for $\beta$-Na$_{0.20}$HfNCl in Ref.\onlinecite{Weht-band}.}}\\[-0.4mm]
\end{tabular}
\end{center}
\end{table}

\subsection{Phonon Spectra}\label{sec:phonon}
\begin{figure*}[t!]
\begin{center}
\includegraphics[scale=0.415]{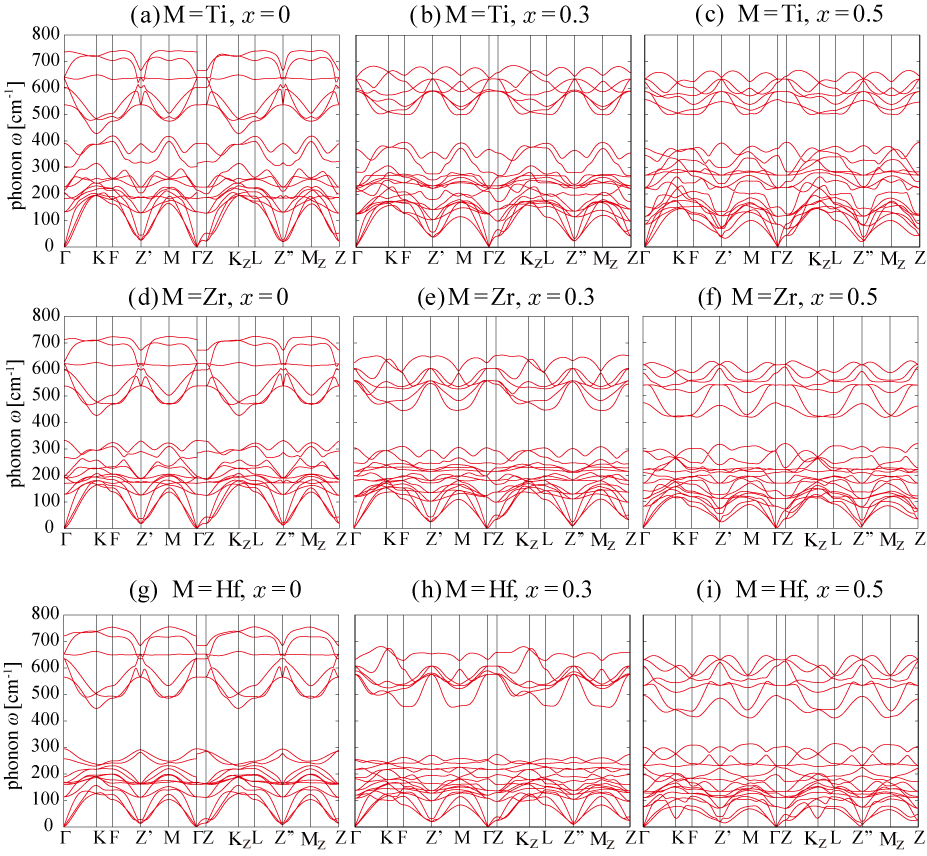}
\caption{(Color online) Phonon dispersions of $\beta$-Li$_{x}$MNCl ($x$=0, 0.3, and 0.5; M=Ti, Zr, and Hf). The acoustic sum rule\cite{Pick-ASR} was applied for the $\Gamma$-point modes.}
\label{fig:phonon}
\end{center}
\end{figure*}

We next show the phonon dispersions in Fig.~\ref{fig:phonon}. The doping (material) dependence can be seen by comparing the panels from left to right (from top to bottom). The total spectrum consists of two bunches of bands; the high-frequency bands around 500--700~cm$^{-1}$ and the low-frequency bands around 0--400~cm$^{-1}$. The high-frequency bands mainly originate from the vibrations of nitrogen atom, while the low-frequency ones are formed by the vibrations of lithium, chlorine, and transition-metal atoms. For the high-frequency bands, Li doping causes appreciable band narrowing and a red shift. On the other hand, for the low-frequency bands, carrier doping basically does not change the overall profile so drastically, but some characteristic change is observed for the Hf compounds; along the Z--K$_{{\rm Z}}$ line, significant mode softening near 0--100~cm$^{-1}$ occurs.

\begin{figure*}[t!]
\begin{center}
\includegraphics[scale=0.58]{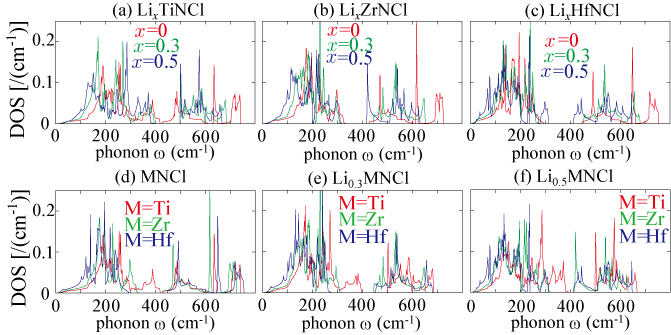}
\caption{(Color online) (a)--(c): Doping dependence of phonon density of states of $\beta$-Li$_{x}$MNCl for each transition metal. (d)--(f): Transition-metal dependence of phonon density of states in $\beta$-Li$_{x}$MNCl with a fixed doping ratio. The density of states are calculated with the tetrahedron integration method\cite{tetrahedra} for the phonon branches interpolated with a 32$\times$32$\times$32  fine mesh.}
\label{fig:phdos}
\end{center}
\end{figure*}

Figure~\ref{fig:phdos} displays the phonon DOS, where the upper three panels describe the doping dependence in the same material and the lower three panels compare the results of the three materials with the same doping rate. The figures show that the carrier doping and the substitution of light elements with heavy elements lead to a red shift of the phonon DOS, especially for the low-frequency range of 0--200~cm$^{-1}$. The origin is explained as follows: For the effect of the carrier doping, $N(0)$ becomes larger (Table~\ref{tab:metaltable}), so that the electronic polarization is enhanced. Thus electronic screening works more effectively and weakens the atomic interaction. On the other hand, heavier atoms vibrate more slowly, so that they generally have lower phonon frequencies.

\subsection{Electron-Phonon Coupling}

\begin{figure*}[t!]
\begin{center}
\includegraphics[scale=0.45]{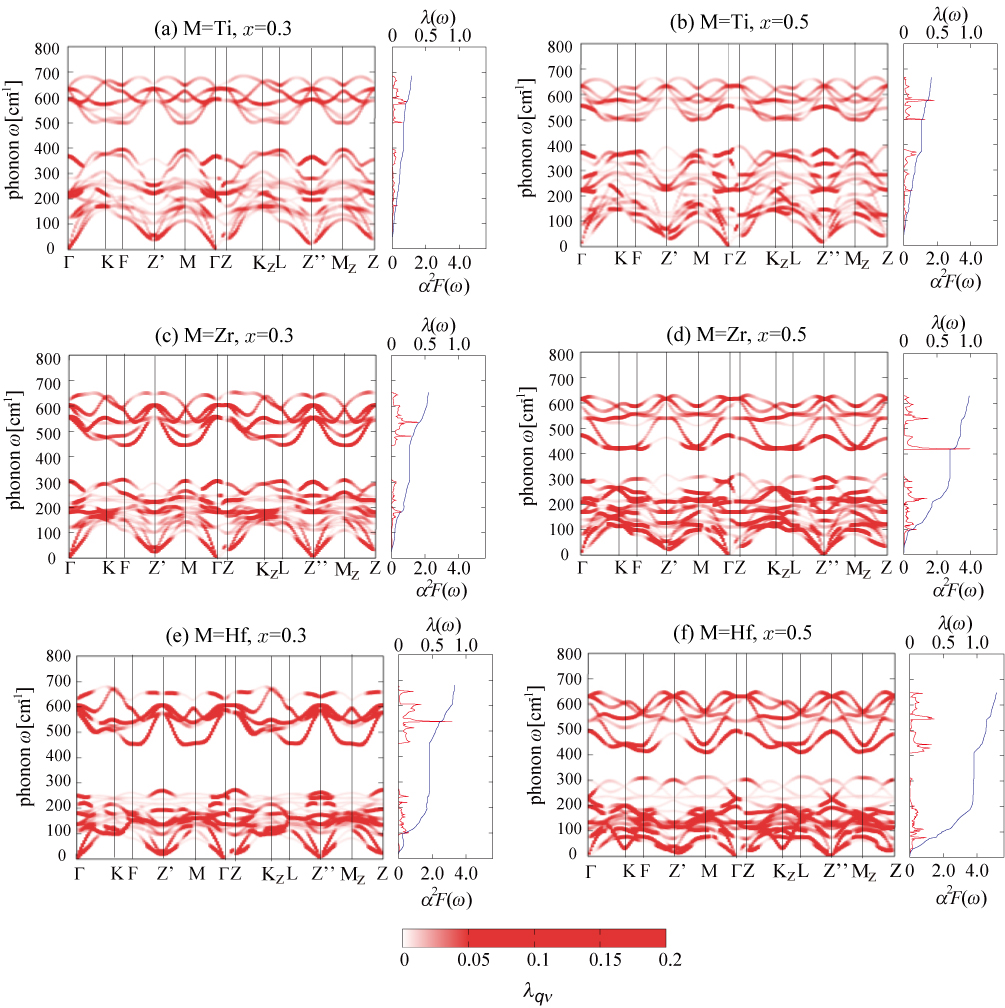}
\caption{(Color online) Electron-phonon couplings for $\beta$-Li$_{x}$MNCl ($x$=0.3 and 0.5; M=Ti, Zr, and Hf); the mode-dependent electron-phonon coupling coefficient $\lambda_{{\bf q}\nu}$ in Eq.~(\ref{lqn}) (left panel), Eliashberg spectral function $\alpha^{2}F(\omega)$ in Eq.~(\ref{a2F}) (right panel), and frequency-dependent coupling coefficient $\lambda(\omega)$ in Eq.~(\ref{lw}) [(blue) solid curves in the right panel]. The Eliashberg function is calculated with the tetrahedron integration method\cite{tetrahedra} for the phonon branches and linewidths interpolated with a 32$\times$32$\times$32 fine mesh.}
\label{fig:elph}
\end{center}
\end{figure*}

\begin{figure*}[t!]
\begin{center}
\includegraphics[scale=0.45]{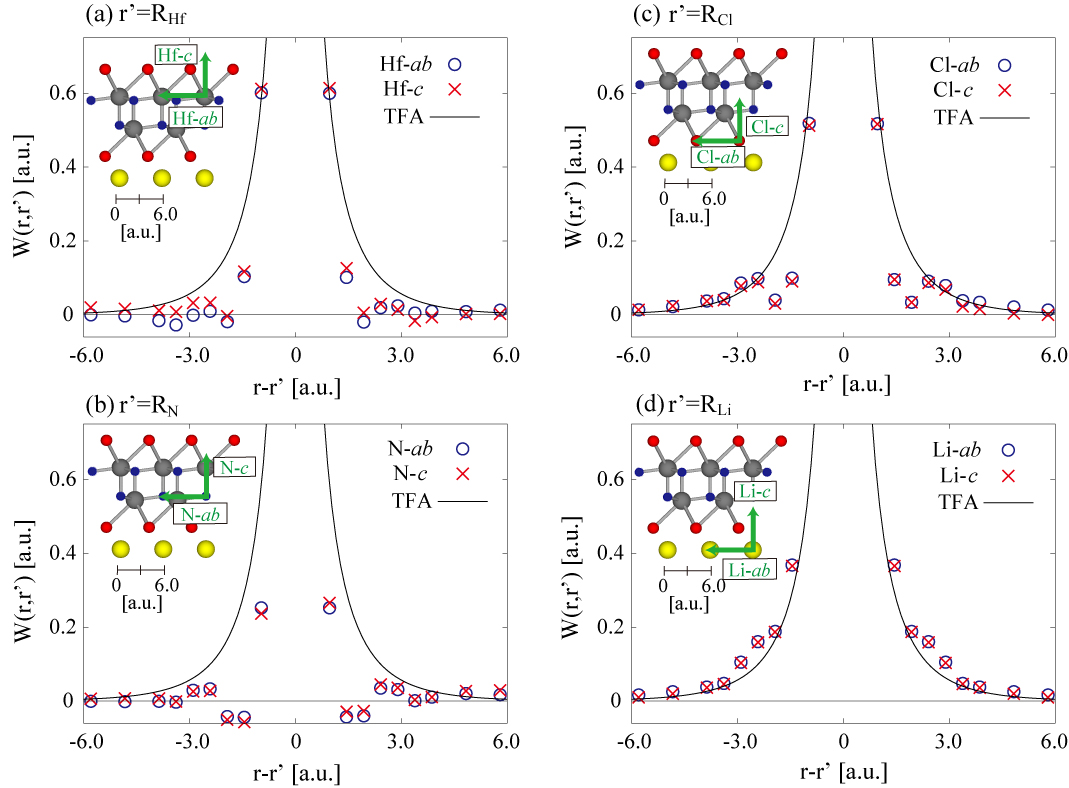}
\caption{(Color online) Our calculated RPA screened electronic interaction $W({\bf r}, {\bf r}')$ in Eq. (\ref{eq:W-RPA-Fourier}) for $\beta$-Li$_{0.3}$HfNCl. The results are shown for the four different origins ${\bf r'}$; ${\bf r'}$=${\bf R}_{\rm Hf}$ in the panel (a), ${\bf r'}$=${\bf R}_{\rm N}$ in (b), ${\bf r'}$=${\bf R}_{\rm Cl}$ in (c), and  ${\bf r'}$=${\bf R}_{\rm Li}$ in (d), where ${\bf R}_{\rm X}$ is the atomic coordinate of the species ${\rm X}$. The calculations were performed along the two directions; one is the direction along the interlayer $c$ axis, plotted as red crosses, and the other is taken to the parallel direction to an Hf-N bond in the $ab$ plane as blue open circles. The geometrical configuration from the side view about the origin and the calculated direction is given in the inset. The figures also include the interaction based on TFA in Eq. (\ref{eq:W-TF-Fourier}) (solid lines). We do not show the values for $|{\bf r}$$-$${\bf r'}|$$\lesssim$$1$ (a.u.), because this regime is beyond the real-space resolution with the present plane-wave cutoff employed in the screened-interaction calculation (12.8 Ry).~\cite{comment-Wr}}
\label{fig:el-el}
\end{center}
\end{figure*}

To estimate the strength of the electron-phonon coupling, we calculated the mode-dependent $\lambda_{{\bf q}\nu}$ as  
\begin{eqnarray}
\lambda_{{\bf q}\nu}=\frac{1}{\pi N(0)} \frac{\gamma_{{\bf q}\nu}}{\omega_{{\bf q}\nu}^{2}}.
\label{lqn}
\end{eqnarray}
Note that $\lambda_{{\bf q}\nu}$ is finite for metals because $\gamma_{{\bf q}\nu}$ in the numerator is defined via the Fermi-surface integral [see Eq.~(\ref{eq:gamma-def})]. The results are displayed in Fig.~\ref{fig:elph}. We also plot in the right side the Eliashberg function $\alpha^{2}F(\omega)$ in Eq.~(\ref{a2F}) and accumulated-frequency function $\lambda(\omega)$ defined as
\begin{eqnarray}
\lambda(\omega)=2\int_{0}^{\omega}d\omega'\frac{\alpha^{2}F(\omega')}{\omega'}.
\label{lw}
\end{eqnarray}
The strength of the electron-phonon coupling can be measured by $\lambda(\omega)$ and the MAD parameter $\lambda$ in Eq.~(\ref{eq:lambda-def}) is defined as the $\omega$$\rightarrow$$\infty$ limit of $\lambda(\omega)$. We see that carrier doping enhances the electron-phonon coupling (compare the left and right panels). Similarly, the electron-phonon coupling is strong in the systems with heavier transition-metal elements (compare the top, middle, and bottom panels).

Table~\ref{tab:metaltable} lists our calculated $\lambda$ in Eq.~(\ref{eq:lambda-def}) to quantify the electron-phonon coupling of each material and each doping rate. In terms of $\lambda$, we expect high $T_{c}$ for heavier transition-metal compounds in the high-doping regime. However, on the other hand, the averaged phonon energy $\omega_{{\rm ln}}$ in Eq.~(\ref{eq:omegaln-def}), listed in the table, exhibits an opposite trend to $\lambda$. In fact, as represented in the MAD formula, the subtle balance of $\lambda$ and $\omega_{{\rm ln}}$ determines $T_{\rm c}$.

\subsection{Screened electron-electron interaction}\label{sec:el-el}
\begin{figure}[t!]
\begin{center}
\includegraphics[scale=0.45]{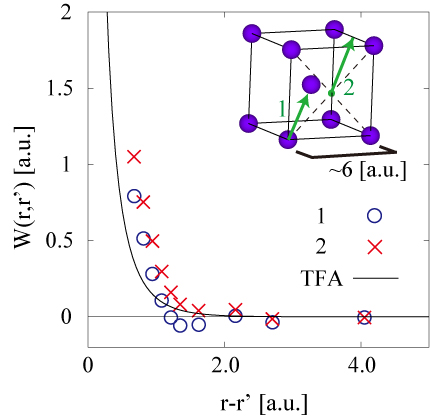}
\caption{(Color online) Our calculated RPA screened electronic interaction $W({\bf r}, {\bf r}')$ in Eq.~(\ref{eq:W-RPA-Fourier}) for Nb. The results are shown for the two different origins ${\bf r'}$; one is set to the Nb atom (denoted by ``1") and the other is on the center of the (100) surface (``2"). The calculations were performed along the (111) direction. The geometrical configuration is schematically given in the inset, where purple spheres represent the Nb atoms and the direction taken in the $W({\bf r}, {\bf r}')$ calculation is indicated by green arrows. The figures also include the interaction based on TFA in Eq.~(\ref{eq:W-TF-Fourier}) (solid lines).}
\label{fig:el-el-Nb}
\end{center}
\end{figure}
We next consider the RPA screened electronic interaction $W({\bf r}, {\bf r}')$ [Eq. (\ref{eq:W-RPA-Fourier})], which is related to $\mathcal{K}^{\rm el}$ by Eq.~(\ref{eq:Coulomb-def}). We calculated it with various settings of the origin and direction. We chose the four points as the origin ${\bf r}'$: The transition-metal, nitrogen, chlorine, and lithium sites. For the calculated directions, we selected the following two: One is the direction along the interlayer {\em c} axis and the other is taken to be parallel to an M-N bond in the {\em ab} plane. We show in Fig.~\ref{fig:el-el} our calculated $W({\bf r}, {\bf r}')$ for $\beta$-Li$_{0.3}$HfNCl. The atomic geometry is depicted in the inset, where the origin and calculated direction are superposed as the green arrow. From the comparison among the four panels, we see significantly weak interactions around the Hf and N atoms building the block layer compared to the TFA one [solid lines, Eq.~(\ref{eq:W-TF-Fourier})], especially for the long-range part ($\geq$2 a.u.). In contrast, the RPA interaction around the Cl and Li atoms seems to follow the TFA interaction. The strong screening around the Hf and N atoms is because of the electronic states near the conduction-band bottom formed by the Hf $5d$ and N $2p$ orbitals.~\cite{Felser-band, Weht-band} In the vicinity of ${\bf r'}$, the interaction around the Ni site  seems to be smaller than that around the Hf one. This might be a consequence of the difference in the electronegativity (1.30 for Hf and 3.04 for N), but it should be noticed that this region is at the border of the resolution limit in the real space with the plane-wave cutoff of 12.8 Ry for screened Coulomb interaction in Eq.~(\ref{WRPAGG}).~\cite{comment-Wr}
The dependence of the RPA screened Coulomb interaction on the atomic species at the origin ${\bf r}'$ and the difference in the short-range part in the RPA and TFA interactions indicates that inhomogeneity not described by TFA is essential for this system. To show this point clearer, we calculated $W({\bf r}, {\bf r}')$ for a simple metal Nb, shown in Fig.~\ref{fig:el-el-Nb}. The ${\bf r}'$ points were set to ($0, 0, 0$) (Nb site,``1" in the inset) and (1/2, 1/2, 0) (``2"). The calculation was performed along the (111) direction (green arrow). We see that the $W({\bf r}, {\bf r}')$ hardly depend on the origin and the calculated directions, which is ascribed to a nearly uniform electronic density in the system. The interaction value itself is, contrary to the $\beta$-Li$_{0.3}$HfNCl case, larger than the TFA one.

The above-mentioned aspects of $W({\bf r}, {\bf r}')$ for $\beta$-Li$_{0.3}$HfNCl are common in essence to the other $\beta$-Li$_{x}$MNCl compounds. On the other hand, as the transition-metal element is replaced by a lighter one, we find that the reduction of the RPA screened interaction around the transition-metal site from the TFA one becomes more appreciable. This may be due to the following fact: When transition metal element is replaced from 5$d$ to 4$d$ or 3$d$, in general, the ionic radius becomes smaller and the electronegativity becomes larger.

Next, to see the strength of the effective interaction between the electrons forming each Cooper pair, we calculated $\mu$ following Eq.~(\ref{eq:mu-def}), with $\mathcal{K}^{\rm el,TFA}$ in Eq.~(\ref{eq:K-el-TF}) and $\mathcal{K}^{\rm el,RPA}$ in Eq.~(\ref{eq:K-el-RPA}). The results for each material and each doping rate are given in Table~\ref{tab:metaltable}. As basic trends, we see that (i) $\mu_{{\rm RPA}}$ is considerably small compared to $\mu_{{\rm TFA}}$ (ii) the large doping rate gives the large $\mu$ value, and (iii) in the order of Hf, Zr, and Ti compounds (i.e., as the transition metal atom becomes lighter), the resulting $\mu$ becomes larger. The trend (i) results from the damped RPA interaction around the nitrogen and transition-metal atoms. The trends (ii) and (iii) originate from the enhancement in the $N(0)$ factor; although the averaged interaction $\langle\!\langle \mathcal{K}^{\rm el} \rangle\!\rangle_{\rm FS}$ itself in Eq.~(\ref{KelFS}) is weakened by the doping or almost unchanged with the substitution from heavier to lighter element, the increase of $N(0)$ (see Fig.~\ref{fig:bands-dos} and Table~\ref{tab:metaltable}) dominates the overall trend in $\mu$ over the $\langle\!\langle \mathcal{K}^{\rm el} \rangle\!\rangle_{\rm FS}$ factor. We also give in the table the renormalized value $\mu^{*}$. The trends of $\mu^{*}$ basically follow those of $\mu$. However, as is well appreciated, the renormalization formula in Eq.~(\ref{eq:renorm}) includes an empirical parameter $E_{{\rm el}}$ and the setting detail of this parameter strongly affect the final value. Hence, the estimate itself has little quantitative reliability.

\subsection{Transition temperatures}
\begin{figure}[h!]
\begin{center}
\includegraphics[scale=0.51]{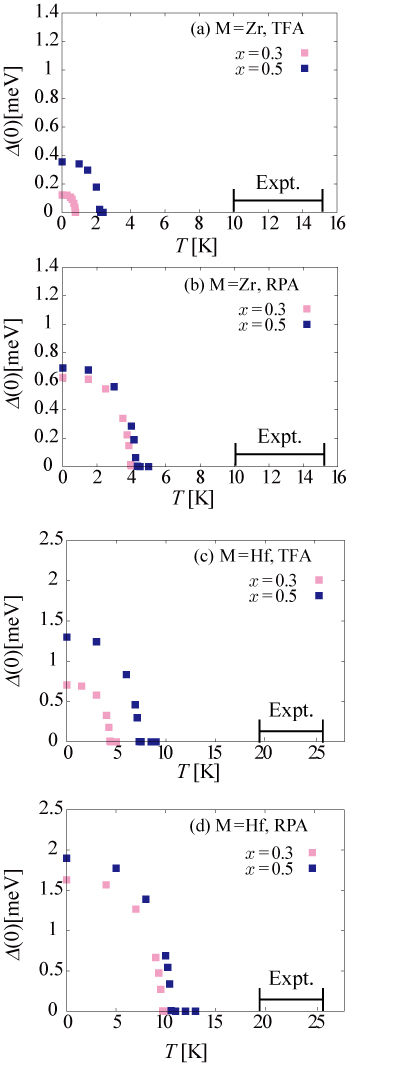}
\caption{(Color online) Temperature dependence of the gap functions for $\beta$-Li$_{x}$MNCl ($x$=0.3 and 0.5; M=Zr and Hf). The panels (a) and (c) are the results with the Thomas-Fermi approximation for the electronic interaction kernel $\mathcal{K}^{\rm el,TFA}$ in Eq.~(\ref{eq:K-el-TF}), and (b) and (d) correspond to the random-phase approximation for $\mathcal{K}^{\rm el,RPA}$ in Eq.~(\ref{eq:K-el-RPA}). The experimental $T_{\rm c}$ ranges are also shown. See the caption of Table \ref{tab:Tc}.}
\label{fig:T-gap}
\end{center}
\end{figure}

We solved the SCDFT equation [Eq.~(\ref{eq:gap-eq}) or (\ref{eq:gap-eq-aux})] with the constructed exchange-correlation kernels $\mathcal{Z}^{\rm ph}$, $\mathcal{K}^{\rm ph}$, and $\mathcal{K}^{\rm el}$ from first principles. Figure~\ref{fig:T-gap} plots our calculated superconducting gap as a function of temperature. The upper two panels (a) and (b) describe results for the Zr compounds and the lower two (c) and (d) describe those for the Hf compounds. The panels (a) and (c) [(b) and (d)] are the results where TFA (RPA) is adopted for $\mathcal{K}^{\rm el}$. The pink (lighter) and blue (darker) squares in the figures represent the results for $x$=0.3 and 0.5, respectively, and the horizontal bars indicate the experimental $T_c$ range.

We find two discernible differences between TFA and RPA, which are common in both the Zr and Hf compounds. First, unlike the results for Al and Nb in Sec. III, the estimated $T_{\rm c}$ is higher in RPA than TFA. This is ascribed to the weaker RPA electronic interaction than the TFA one (Sec.~\ref{sec:el-el}). Second, while both TFA and RPA predict that an electron doping raises $T_{\rm c}$, the increasing trend of $T_{\rm c}$ in RPA is less noticeable than that in TFA. This is because that the weakening of the RPA interaction in the MN layers (Fig.~\ref{fig:el-el}) compared to the TFA one becomes more appreciable in the low-doping regime than the high-doping one (as seen in Table \ref{tab:metaltable}), thus leading to the relatively higher RPA $T_{\rm c}$ than the TFA $T_{\rm c}$ at $x$$=$$0.3$. The observed differences between TFA and RPA highlight a significance of the effects beyond TFA in $\beta$-Li$_{x}$MNCl.

Importantly, our estimated $T_{\rm c}$ is a few times lower than the experimental values, being in a sharp contrast with the case of simple metals for which SCDFT successfully reproduces experimental $T_{\rm c}$. The disagreement observed for $\beta$-Li$_x$MNCl suggests that the present exchange-correlation kernels $\mathcal{Z}^{\rm ph}$, $\mathcal{K}^{\rm ph}$, and $\mathcal{K}^{\rm el}$ is missing something crucial to describe the superconductivity in this system.
\begin{table}[t!]
\caption[t]
{Our calculated superconducting transition temperatures (K) with density functional theory for superconductors and the McMillan-Allen-Dynes formula in Eq.~(\ref{eq:McMillan}), and the parameter $\mu^{\ast}$ given so that Eq.~(\ref{eq:McMillan}) reproduce $T_{\rm c}^{{\rm SCDFT\mathchar`-RPA}}$.}
\begin{center}
\label{tab:Tc}
{\tabcolsep = 1mm
\begin{tabular}{lcccccc} \toprule[2pt]
M & \multicolumn{2}{c}{Ti} & \multicolumn{2}{c}{Zr} & \multicolumn{2}{c}{Hf} \\
\midrule[1pt]
\textit{x} & 0.3 & 0.5 & 0.3 & 0.5 & 0.3 & 0.5 \\
\midrule[0.5pt]
$T_{\rm c}^{{\rm SCDFT\mathchar`-}\mathcal{K}^{\rm el}=0}$ &6.3 &7.4  &17.8  &33.0  &28.6  &36.6  \\
$T_{\rm c}^{\rm SCDFT\mathchar`-TFA}$  & -- & -- &\ph 0.8 &\ph 2.2 &\ph 4.4 &\ph 7.2 \\
$T_{\rm c}^{\rm SCDFT\mathchar`-RPA}$  & -- & -- &\ph 3.9 &\ph 4.3 &\ph 9.6 &10.5 \\
\midrule[0.2pt]
$T_{\rm c}^{\rm MAD}(\mu^{\ast}$$=$$0)$ &3.6 &10.1 &18.8 &31.2 &27.1 &26.9 \\
$T_{\rm c}^{\rm MAD}(\mu^{\ast}_{\rm TFA})$ & -- & -- &\ph 2.1 &\ph 9.7  &\ph 7.2  &14.6 \\
$T_{\rm c}^{\rm MAD}(\mu^{\ast}_{\rm RPA})$ & $<$10$^{-3}$ & $<$10$^{-3}$ &\ph 6.2 &11.5 &14.8 &16.6 \\
\midrule[0.2pt]
$T_{\rm c}^{\rm expt}$ & -- & -- & \multicolumn{2}{c}{10.0--15.2$^{a}$} & \multicolumn{2}{c}{19.0--25.5$^{b}$} \\
\midrule[0.2pt]
$\mu^{\ast}_{\rm SCDFT\mathchar`-RPA}$  &\ \ --\ \ \ &\ \ --\ \ \ &0.144 &0.296 &0.170 &0.242 \\
\midrule[2pt]
\multicolumn{7}{l}{{\scriptsize $^{a}$Ref.~\onlinecite{Taguchi-doping} ($\beta$-Li$_{x}$ZrNCl); Ref.~\onlinecite{Kasahara2009} ($\beta$-Li$_{x}$ZrNCl);}}\\[-0.4mm]
\multicolumn{7}{l}{{\scriptsize Ref.~\onlinecite{Hotehama-molecular} [$\beta$-Li$_{x}$M$_{y}$ZrNCl; M= tetrahydrofuran (THF),}}\\[-0.4mm]
\multicolumn{7}{l}{\hspace{30pt}{\scriptsize propylene carbonate (PC)];}}\\[-0.4mm]
\multicolumn{7}{l}{{\scriptsize Ref.~\onlinecite{Kawaji} ($\beta$-A$_{x}$ZrNCl; A=Li, Na, K);}}\\[-0.4mm]
\multicolumn{7}{l}{{\scriptsize Ref.~\onlinecite{Kasahara-molecular} [$\beta$-Li$_{x}$M$_{y}$ZrNCl; M=(N, N)-dimethylformamide,}}\\[-0.4mm]
\multicolumn{7}{l}{\hspace{30pt}{\scriptsize diemthylsulfoxide].}}\\[-0.4mm]
\multicolumn{7}{l}{{\scriptsize $^{b}$Ref.~\onlinecite{Yamanaka-nat} [$\beta$-Li$_{0.48}$(THF)$_{y}$HfNCl]; Ref.~\onlinecite{Shamoto2004} ($\beta$-Na$_{x}$HfNCl);}}\\[-0.4mm]
\multicolumn{7}{l}{{\scriptsize Ref.~\onlinecite{Takano-molecular} ($\beta$-Li$_{x}$M$_{y}$HfNCl; M=NH$_{3}$, THF, PC);}}\\[-0.4mm]
\multicolumn{7}{l}{{\scriptsize  Ref.~\onlinecite{Hotehama-molecular} ($\beta$-A$_{x}$M$_{y}$HfNCl; A= Li, Na; M= THF, PC).}}\\[-0.4mm]
\end{tabular}
}
\end{center}
\end{table}

We summarize the SCDFT results in Table~\ref{tab:Tc}. We compare three $T_{\rm c}$s obtained from different treatments of the Coulomb interaction effect;
$T_{\rm c}^{{\rm SCDFT\mathchar`-}\mathcal{K}^{\rm el}=0}$ for which $\mathcal{K}^{\rm el}$ is neglected in the gap function, $T_{\rm c}^{\rm SCDFT\mathchar`-TFA}$ calculated with TFA, and $T_{\rm c}^{\rm SCDFT\mathchar`-RPA}$ obtained by using RPA. For the Ti compounds, we found no superconducting solution in calculations with $\mathcal{K}^{\rm el}$. We also list $T_c$ estimated by the MAD formula. The trend is basically the same as that of SCDFT. While the MAD gives higher $T_c$, it should be noted that the MAD formula contains the empirical parameter. As a reference, we estimated $\mu^{*}=\mu^{*}_{\rm SCDFT\mathchar`-RPA}$ which reproduces our $T_c^{\rm SCDFT\mathchar`-RPA}$ for the Zr and Hf compounds and list them at the bottom of the table. While the empirical formula [Eq.~(\ref{eq:renorm})] gives 0.113--0.196 (see Table~\ref{tab:metaltable}), $\mu^{*}_{\rm SCDFT\mathchar`-RPA}$ is 0.144--0.296, which is much larger than the standard values ($\sim$$0.10$) used in literatures.\cite{Carbotte} These results suggest that suppression of the Coulomb interaction by screening and renormalization in the $\beta$-MNCl is not so strong.

\begin{figure}[b!]
\begin{center}
\includegraphics[scale=0.8]{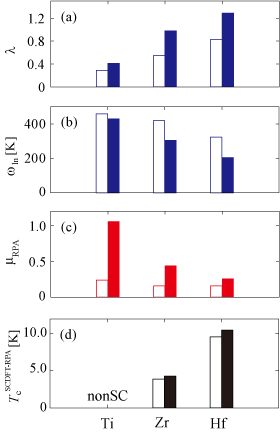}
\caption{(Color online) Transition-metal and doping dependence of the {\it ab initio} McMillan-Allen-Dynes parameters; $\lambda$ in the panel (a), $\omega_{\rm ln}$ in (b), $\mu_{\rm RPA}$ in (c), and $T^{\rm SCDFT\mathchar`-RPA}_{\rm c}$ in (d). See also Tables~\ref{tab:metaltable} and \ref{tab:Tc}. Empty and filled boxes represent the values for the $x$=0.3 and $x$=0.5 compounds, respectively. For the panels (a)--(c), the boxes drawn in blue (red) indicate that the corresponding parameter contributes to raising (lowering)  $T_{\rm c}$.}
\label{fig:params-metal}
\end{center}
\end{figure}
\section{discussion}\label{sec:discussion}
In discussing overall aspect of the present SCDFT calculation, an analysis with the parameters in the MAD formula gives a simple insight because each parameter represents an average of the matrix quantity in SCDFT. We summarize in Fig.~\ref{fig:params-metal} these parameters; $\lambda$ in the panel (a), $\omega_{{\rm ln}}$ in (b), $\mu_{{\rm RPA}}$ in (c), and $T_{c}^{\rm SCDFT\mathchar`-RPA}$ in (d). The figure compares the three compounds (Ti, Zr, and Hf) and the empty and solid bars represent the results for $x$=0.3 and 0.5, respectively. While the material dependence of $T_c$ basically follows the $\lambda$ trend, the quantitative aspect of $T_c$ is determined as a balance in $\lambda$, $\omega_{\rm ln}$, and $\mu$. Because of the rather large $\mu_{\rm RPA}$, the Ti compounds do not exhibit superconductivity. The doping dependence of $T_{\rm c}$ also follows the $\lambda$ trend. However, because of the decreasing (increasing) trend of $\omega_{\rm ln}$ ($\mu_{\rm RPA}$) upon the doping, the $T_c$ difference in $x$=0.3 and 0.5 is not substantial.
\begin{figure}[t]
\begin{center}
\includegraphics[scale=0.68]{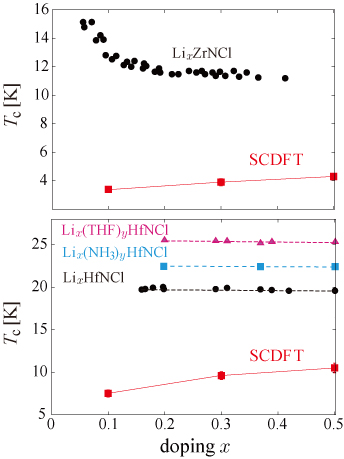}
\caption{(Color online) Doping dependence of transition temperatures derived from SCDFT in $\beta$-Li$_{x}$ZrNCl (upper panel) and $\beta$-Li$_{x}$HfNCl (lower panel). The errors due to the random sampling in the SCDFT calculation are also shown with the vertical bars. Experimental transition temperatures are taken from Ref.\onlinecite{Taguchi-doping} for the Zr compound and Ref.\onlinecite{Takano-molecular} for the Hf one. The notation THF stands for tetrahydrofuran. The dashed lines are guides to the eyes.}
\label{fig:Tc-dope}
\end{center}
\end{figure}
Besides the overall trend of each factor, there is a notable quantitative disagreement in $T_{\rm c}$ between the theory and the experiment. Since the SCDFT assures a high accuracy of $T_{\rm c}$ for a wide class of phonon-mediated superconductors, this disagreement is remarkable and should be seriously and carefully examined. To study this point in more detail, we calculated the doping dependence of $T_c$ for lower doping regime down to $x$$=$$0.1$. In the experiments,\cite{Taguchi-doping, Takano-molecular} $T_{\rm c}$ is known to rapidly increase as $x$ is reduced down to $\sim$$0.05$ for the Zr compounds, while it barely changes for $x$$>$$0.1$ for the Hf compounds. In order to examine whether this behavior is reproduced theoretically, we estimated $T_{\rm c}$ for $x$$=$$0.1$ by SCDFT. In this calculation, the RPA electronic kernel in Eq.~(\ref{eq:K-el-RPA}) is made with the rigid-band-shift approximation to the electronic state for $x$$=$$0.3$. For $\alpha^{2}F$, we employed that of $x$$=$$0.3$. This is justified from the fact that the DOS at $E_{\rm F}$ for $x$$=$$0.1$ is nearly the same as that for $x$$=$$0.3$ (see Fig.~\ref{fig:bands-dos}). The results are plotted in Fig.~\ref{fig:Tc-dope} and compared with the experiments. We see that the resulting $T_{\rm c}$ lowers for small $x$, which is an opposite trend to the experiments. The doping dependence by SCDFT is reasonable because the screening due to the virtual states in the flat-band region just above $E_{\rm F}$ becomes less effective for small $x$ (as $E_{\rm F}$ is apart from this flat band). Indeed, the bare $\mu$ is estimated to be $0.203$ for the Zr compound and $0.237$ for the Hf compound, which are respectively larger than those for $x$$=$$0.3$ ($0.161$ for both of the Zr and Hf compounds). 

The present SCDFT is based on the standard ME theory, in which one neglects the momentum dependence of the quantities. Also, the screened Coulomb interaction is described with RPA and the frequency dependence is ignored. The present discernible difference in $T_{\rm c}$ or in its doping dependence between the theory and experiment suggest the importance of the effect not included in the analysis in the ME level. In fact, in the optical experiment,\cite{Takano-optical} $\beta$-Li$_x$ZrNCl is known to have the plasma edge in the low-energy Drude band less than $\sim$0.1 eV, exhibiting a redshift with the lowering doping amount. Since the presence of the low-energy plasma edge affects the screened Coulomb interaction in low-frequency regime, it can be relevant to the low-energy properties of the present system as proposed earlier in Ref.~\onlinecite{Bill-rapid2002}. Meanwhile, there remains a possibility that correction in the Kohn-Sham orbitals improves the present error; for example, the use of the hybrid functional in deformation-potential calculation seems to lead to the increase of the electron-phonon coupling.\cite{Yin-Kotliar} It will be interesting that the present SCDFT calculation is performed with such hybrid functional to study the degree of the improvement.

\section{Summary and outlook}\label{sec:sum}
We have performed a comprehensive SCDFT analysis for doped $\beta$-MNCl, where M=Ti, Zr and Hf, to study the effects described within the Migdal-Eliashberg theory without introducing any adjustable parameters. On the basis of the \textit{ab initio} electronic band structure, phonon spectrum, electron-phonon coupling, and electron-electron interaction in the RPA level, we estimated superconducting $T_{\rm c}$. The resulting $T_{\rm c}$ values are $\lesssim$4 K for the Zr compounds and $\lesssim$10 K for the Hf ones, whereas the Ti compounds do not show superconductivity because of the large electronic interaction.  While the dependence of $T_{\rm c}$ on the transition-metal species is consistent with the experiments, our calculated $T_{\rm c}$ values are less than a half of the experimental values (10--15 K for the Zr case and 19--26 K for the Hf one) and the theoretical doping dependence of $T_{\rm c}$ is opposite to that of the experiment. Although we do not rule out some possibilities of improvement by corrections to one-particle states as the starting point, the revealed discrepancy between the theory and experiment, as is known in the case of the unconventional superconductors, appeals the relevance of something ignored in the standard Migdal-Eliashberg theory to the superconducting mechanism of the doped $\beta$-MNCl. This remains an open question. 
\\
\\
\begin{acknowledgments}
We thank Mitsuaki Kawamura for stimulating discussions. This work was supported by Grants-in-Aid for Scientific Research (No.~22740215, 22104010, 23110708, 23340095, 19051016), Funding Program for World-Leading Innovative R\&D on Science and Technology (FIRST program) on ``Quantum Science on Strong Correlation'', JST-PRESTO and the Next Generation Super Computing Project and Nanoscience Program from MEXT, Japan.
\end{acknowledgments}

\begin{appendix}
\section{$\Gamma$-point Phonons}\label{app:Gammaphon}
To check the reliability for phonon properties, we calculated vibrational frequencies at the $\Gamma$ point and compared with previous theoretical and experimental results. Tables~\ref{tab:Raman-nondoped} and \ref{tab:IR-nondoped} list the Raman and infrared frequencies for the undoped system, respectively. Also, Tables~\ref{tab:Raman-doped} and \ref{tab:IR-doped} are the results for doped samples. For both of the undoped and doped samples, our calculations reasonably reproduces the experimental results, as well as the previous theoretical results. 
\\
\\
\\
\\
\\
\\
\begin{table}[h]
\caption[h!]
{Our calculated frequencies (cm$^{-1}$) of the Raman-active modes for undoped $\beta$-MNCl.}
\begin{center}
\label{tab:Raman-nondoped}
{\scriptsize 
\begin{tabular}{lccccccccc} \toprule[2pt]
 M& Ti& &\multicolumn{3}{c}{Zr} & & \multicolumn{3}{c}{Hf} \\
 \cline{2-2}\cline{4-6}\cline{8-10}
  & present & & present & expt.$^{a}$ & 
  theory$^{b}$ 
  & & present & expt.$^{a}$ & theory$^{c}$ 
\\
\midrule[0.5pt]
A$_{1g}$\#1 &603 & &598  &591  &564 & &634 &612 &604 \\
A$_{1g}$\#2 &391 & &333  &331  &312 & &295 &292 &386 \\
A$_{1g}$\#3 &214 & &196  &191  &182 & &169 &163 &191 \\
E$_{g}$\#1  &640 & &622  &605  &620 & &652 &633 & -  \\
E$_{g}$\#2  &227 & &191  &184  &171 & &166 &156 & -  \\
E$_{g}$\#3  &133 & &130  &128  &114 & &113 &110 & -  \\
\midrule[2pt]
\multicolumn{10}{l}{$^{a}$Raman spectroscopy taken from Ref.~\onlinecite{Cros}.}\\
\multicolumn{10}{l}{$^{b}${\it Ab initio} DFT-LDA pseudopotential calculation in Ref.~\onlinecite{Heid-DFPT}.}\\
\multicolumn{10}{l}{$^{c}${\it Ab initio} DFT-LDA pseudopotential calculation in Ref.~\onlinecite{Weht-band}.}\\
\end{tabular}
} 
\end{center}
\end{table}
\begin{table}[b!]
\caption[t]
{Our calculated frequencies (cm$^{-1}$) for infrared-active modes for undoped $\beta$-MNCl. The values in the parentheses are the frequencies of the longitudinal vibrations.}
\begin{center}
\label{tab:IR-nondoped}
{\scriptsize 
\begin{tabular}{lcccccccc} \toprule[2pt]
M & Ti & & \multicolumn{3}{c}{Zr} & & \multicolumn{2}{c}{Hf} \\
 \cline{2-2}\cline{4-6}\cline{8-9}
  & present &
  & present & expt.$^{a}$ & 
  theory$^{b}$ 
  & & present & theory$^{c}$ 
  \\
\midrule[0.5pt]
A$_{2u}$\#1 &638(666) & &630(673) &529  &597(630) & &638(684) &617 \\
A$_{2u}$\#2 &303(317) & &268(286) &-    &257(272) & &257(276) &128 \\
E$_{u}$\#1  &537(736) & &538(713) &666? &547(680) & &565(721) &632 \\
E$_{u}$\#2  &184(206) & &171(198) &165  &155(179) & &159(189) &105 \\
\midrule[2pt]
\multicolumn{9}{l}{$^{a}$Infrared spectroscopy taken from Ref.~\onlinecite{Adelmann}.}\\
\multicolumn{9}{l}{$^{b}${\it Ab initio} DFT-LDA pseudopotential calculation in Ref.~\onlinecite{Heid-DFPT}.}\\
\multicolumn{9}{l}{$^{c}$Force-constant-model calculation reproducing}\\
\multicolumn{9}{l}{the Raman spectra in Ref.~\onlinecite{Cros}.}\\
\end{tabular}
} 
\end{center}
\end{table}
\begin{table}[t!]
\caption[b!]
{Our calculated frequencies (cm$^{-1}$) of the Raman-active modes for doped $\beta$-Li$_{x}$MNCl.}
\begin{center}
\label{tab:Raman-doped}
{\scriptsize 
\begin{tabular}{lccccccccccc} \toprule[2pt]
M & \multicolumn{2}{c}{Ti} & & \multicolumn{4}{c}{Zr} & & \multicolumn{3}{c}{Hf} \\ 
 \cline{2-3}\cline{5-8}\cline{10-12}
 & \multicolumn{2}{c}{present} & 
 & \multicolumn{2}{c}{present} & expt.$^{a}$ & 
 theory$^{b}$
 & & \multicolumn{2}{c}{present} & expt.$^{c}$ \\
 \cline{2-3}\cline{5-6}\cline{7-8}\cline{10-11}\cline{12-12}
$x$ & 0.3 & 0.5 & &  0.3 & 0.5 &0.16 &1/6 & & 0.3 & 0.5 & 0.35  \\
\midrule[0.5pt]
A$_{1g}$\#1 &594 &556 & &557 &473 &582 &565--548 & &589 &496 &595 \\
A$_{1g}$\#2 &369 &372 & &304 &291 &322 &303--296 & &250 &228 &282 \\
A$_{1g}$\#3 &210 &223 & &198 &209 &188 &186--217 & &181 &197 &147 \\
E$_{g}$\#1  &634 &633 & &603 &616 &608 &599--633 & &607 &631 &616 \\
E$_{g}$\#2  &222 &224 & &181 &170 &178 &169--165 & &138 &134 &157 \\
E$_{g}$\#3  &125 &116 & &123 &111 &123 &116--131 & &119 &110 &106 \\
\midrule[2pt]
\multicolumn{12}{l}{$^{a}$Raman spectroscopy taken from Ref.~\onlinecite{Adelmann}.}\\
\multicolumn{12}{l}{$^{b}${\it Ab initio} DFT-LDA pseudopotential calculations in Ref.~\onlinecite{Heid-DFPT}.}\\
\multicolumn{12}{l}{$^{c}$Raman spectroscopy for $\beta$-Na$_{x}$HfNCl taken from Ref.~\onlinecite{Cros}.}\\
\end{tabular}
} 
\end{center}
\end{table}

\begin{table}[t!]
\caption[t]
{Our calculated frequencies (cm$^{-1}$) for infrared-active modes for doped $\beta$-Li$_{x}$MNCl.}
\begin{center}
\label{tab:IR-doped}
{\scriptsize 
\begin{tabular}{lcccccccccc} \toprule[2pt]
M & \multicolumn{2}{c}{Ti} & & \multicolumn{3}{c}{Zr} & &  \multicolumn{3}{c}{Hf} \\
 \cline{2-3}\cline{5-7}\cline{9-11}
 & \multicolumn{2}{c}{present} &
 & \multicolumn{2}{c}{present} &
 theory$^{a}$ 
 & & \multicolumn{2}{c}{present} &
 theory$^{b}$
 \\
 \cline{2-3}\cline{5-6}\cline{7-7}\cline{9-10}\cline{11-11}
$x$ & 0.3 & 0.5 & & 0.3 & 0.5 &1/6 & & 0.3 & 0.5 &0.5 \\
\midrule[0.5pt]
A$_{2u}$\#1 &638 &622 & &625 &585 &612--583 & &632 &606 &596 \\
A$_{2u}$\#2 &283 &346 & &247 &291 &314 & &254 &300 &108 \\
A$_{2u}$\#3 &223 &152 & &198 &142 &235--184 & &186 &128 &183 \\
E$_{u}$\#1  &588 &587 & &557 &540 &549--603 & &576 &534 &618 \\
E$_{u}$\#2  &231 &271 & &215 &221 &262 & &217 &232 &85  \\
E$_{u}$\#3  &127 &86  & &119 &83  &148--113 & &108 &77  &146 \\
\midrule[2pt]
\multicolumn{11}{l}{$^{a}${\it Ab initio} DFT-LDA pseudopotential calculations}\\
\multicolumn{11}{l}{in Ref.~\onlinecite{Heid-DFPT}.}\\
\multicolumn{11}{l}{$^{b}$Force-constant-model calculation reproducing}\\
\multicolumn{11}{l}{ the Raman spectra for $\beta$-Na$_{0.58}$HfNCl in Ref.~\onlinecite{Cros}.}\\
\end{tabular}
} 
\end{center}
\end{table}

\end{appendix}


\clearpage
\begin{thebibliography}{999}
\bibitem{Bednorz-Muller} J. G. Bednorz and K. A. Muller, Z. Phys. B \textbf{64}, 189 (1986). 
\bibitem{Yamanaka1996} S. Yamanaka, H. Kawaji, K. Hotehama, M. Ohashi, Adv. Mat. {\bf 8}, 771 (1996).
\bibitem{Yamanaka-nat} S. Yamanaka, K. Hotehama, and H. Kawaji, Nature \textbf{392}, 580 (1998).
\bibitem{Yamanaka-review} S. Yamanaka, J. Mater. Chem. \textbf{20}, 2922 (2010).
\bibitem{Migdal} A. B. Migdal, Sov. Phys. JETP \textbf{34}, 996 (1958)
\bibitem{Eliashberg} G. M. Eliashberg, Sov. Phys. JETP \textbf{11}, 696 (1960)
\bibitem{Scalapino} D. J. Scalapino, in {\it Superconductivity} edited by R. D. Parks, (Marcel Dekker, New York, 1969) VOLUME 1.
\bibitem{Schrieffer} J. R. Schrieffer,{\it Theory of superconductivity; Revised Printing},  (Westview Press, Colorado, 1971)
\bibitem{Boeri2008} L. Boeri, O. V. Dolgov, and A. A. Golubov, Phys. Rev. Lett. \textbf{101}, 026403 (2008).
\bibitem{Tou-isotope} H. Tou, Y. Maniwa, and S. Yamanaka, Phys. Rev. B \textbf{67}, 100509(R) (2003).
\bibitem{Tou-NMR2010} H. Tou, S. Oshiro, H. Kotegawa, Y. Taguchi, Y. Kishiume, Y. Kasahara, and Y. Iwasa, Physica C \textbf{470}, S658 (2010).
\bibitem{Ekino-tunnel} T. Ekino, T. Takasaki, T. Muranaka, H. Fujii, J. Akimitsu, and S. Yamanaka, Physica B \textbf{328}, 23 (2003).
\bibitem{Takasaki-tunnel2005} T. Takasaki, T. Ekino, H. Fujii, and S. Yamanaka, J. Phys. Soc. Jpn. \textbf{74}, 2586-2591 (2005).
\bibitem{Takasaki-tunnel2010} T. Takasaki, T. Ekino, A. Sugimoto, K. Shohara, S. Yamanaka, and A.M. Gabovich, Euro. Phys. J. B \textbf{73}, 471 (2010).
\bibitem{Tou-suscep} H. Tou, Y. Maniwa, T. Koiwasaki, and S. Yamanaka, Phys. Rev. Lett. \textbf{86}, 5775 (2001).
\bibitem{Taguchi-specheat} Y. Taguchi, M. Hisakabe, and Y. Iwasa, Phys. Rev. Lett. \textbf{94}, 217002 (2005).
\bibitem{Taguchi-isotope} Y. Taguchi, T. Kawabata, T. Takano, A. Kitora, K. Kato, M. Takata, and Y. Iwasa, Phys. Rev. B \textbf{76}, 064508 (2007).
\bibitem{BCS} J. Bardeen, L. N. Cooper, and J. R. Schrieffer, Phys. Rev. \textbf{108}, 1175 (1957).
\bibitem{Taguchi-doping} Y. Taguchi, A. Kitora, and Y. Iwasa, Phys. Rev. Lett. \textbf{97}, 107001 (2006).
\bibitem{Takano-molecular} T. Takano, T. Kishiume, Y. Taguchi, and Y. Iwasa, Phys. Rev. Lett. \textbf{100}, 247005 (2008).
\bibitem{Kasahara2009} Y. Kasahara, T. Kishiume, T. Takano, K. Kobayashi, E. Matsuoka, H. Onodera, K. Kuroki, Y. Taguchi, and Y. Iwasa, Phys. Rev. Lett. \textbf{103}, 077004 (2009).
\bibitem{Tou-NMR2007} H. Tou, M. Sera, Y. Maniwa, and S. Yamanaka, Int. J. Mod. Phys. B \textbf{21}, 3340 (2007).
\bibitem{Kotegawa-JPS2011} H. Kotegawa, unpublished.
\bibitem{Hiraishi} M. Hiraishi, R. Kadono, M. Miyazaki, S. Takeshita, Y. Taguchi, Y. Kasahara, T. Takano, T. Kishiume, and Y. Iwasa, Phys. Rev. B \textbf{81}, 014525 (2010).
\bibitem{Kuroki2008} K. Kuroki, Sci. Tech. Adv. Mater. \textbf{9}, 044202 (2008).
\bibitem{Kuroki2010} K. Kuroki, Phys. Rev. B \textbf{81}, 104502 (2010).
\bibitem{Choi-Kuroki2011} S. Choi, H. Aizawa, and K. Kuroki, J. Phys. Chem. Sol. \textbf{72}, 376 (2011).
\bibitem{Yin-Pickett} Q. Yin, E. R. Ylvisaker, and W. E. Pickett, Phys. Rev. B \textbf{83}, 014509 (2011).
\bibitem{Heid-DFPT} R. Heid and K. P. Bohnen, Phys. Rev. B \textbf{72}, 134527 (2005).
\bibitem{Bergman-Rainer} G. Bergmann and D. Rainer, Z. Phys. \textbf{263}, 59 (1973).
\bibitem{Morel-Anderson} P. Morel and P. W. Anderson, Phys. Rev. \textbf{125}, 1263 (1962); N. N. Bogoliubov, V. V. Tolmachev, and D. V. Shirkov, {\it A New Method in the Theory of Superconductivity} (1958) (translated from Russian: Consultants Bureau, Inc., New York, 1959).
\bibitem{Weht-band} R. Weht, A. Filippetti, and W. E. Pickett, Europhys. Lett. \textbf{48}, 320 (1999).
\bibitem{Gaspari-Gyorffy} G. D. Gaspari and B. L. Gyorffy, Phys. Rev. Lett. \textbf{28}, 801 (1972).
\bibitem{Klein-Pickett} B. M. Klein and W. E. Pickett, in \textit{Superconductivity in d- and f-Band Metals 1982}, edited by
W. Buckel and W. Weber (Kernforschungszentrum Karlsruhe, Karlsruhe) 1982, p. 477
\bibitem{Yin-Kotliar} Z. P. Yin, A. Kutepov, and G. Kotliar, arXiv:1110.5751.
\bibitem{HSE} J. Heyd, G. E. Scuseria, and M. Ernzerhof, J. Chem. Phys. \textbf{118}, 8207 (2003); \textbf{124}, 219906(E) (2006).
\bibitem{McMillan} W. L. McMillan, Phys. Rev. \textbf{167}, 331 (1968).
\bibitem{AllenDynes} P. B. Allen and R. C. Dynes, Phys. Rev. B \textbf{12}, 905 (1975).
\bibitem{Carbotte} J. P. Carbotte, Rev. Mod. Phys. \textbf{62}, 1027 (1990)
\bibitem{Bill-rapid2002} A. Bill, H. Morawitz, and V. Z. Kresin, Phys. Rev. B \textbf{66}, 100501(R) (2002).
\bibitem{Kravtsov} V. E. Kravtsov, arxiv:1109.2515.
\bibitem{Burmistrov} I. S. Burmistrov, I. V. Gornyi, and A. D. Mirlin, Phys. Rev. Lett. \textbf{108}, 017002 (2012).
\bibitem{Lee-Chang-Cohen} K. H. Lee, K. J. Chang, and M. L. Cohen, Phys. Rev. B \textbf{52}, 1425 (1995).
\bibitem{Luders} M. L\"uders, M. A. L. Marques, N. N. Lathiotakis, A. Floris, G. Profeta, L. Fast, A. Continenza, S. Massidda, and E. K. U. Gross, Phys. Rev. B \textbf{72}, 024545 (2005).
\bibitem{Oliveira} L. N. Oliveira, E. K. U. Gross, and W. Kohn, Phys. Rev. Lett. \textbf{60}, 2430 (1988).
\bibitem{Marques-metals} M. A. L. Marques, M. L\"uders, N. N. Lathiotakis, G. Profeta, A. Floris, L. Fast, A. Continenza, E. K. U. Gross, and S. Massidda, Phys. Rev. B \textbf{72}, 024546 (2005).
\bibitem{Floris-MgB2} A. Floris, G. Profeta, N. N. Lathiotakis, M. L\"uders, M. A. L. Marques, C. Franchini, E. K. U. Gross, A. Continenza, and S. Massidda, Phys. Rev. Lett. \textbf{94}, 037004 (2005).
\bibitem{Sanna-CaC6} A. Sanna, G. Profeta, A. Floris, A. Marini, E. K. U. Gross, and S. Massidda, Phys. Rev. B \textbf{75}, 020511(R) (2007).
\bibitem{Profeta-pressure} G. Profeta, C. Franchini, N. N. Lathiotakis, A. Floris, A. Sanna, M. A. L. Marques, M. L\"uders, S. Massidda, E. K. U. Gross, and A. Continenza, Phys. Rev. Lett. \textbf{96}, 047003 (2006).
\bibitem{Bersier-CaBeSi} C. Bersier, A. Floris, A. Sanna, G. Profeta, A. Continenza, E. K. U. Gross, and S. Massidda, Phys. Rev. B \textbf{79}, 104503 (2009).
\bibitem{Massidda} S. Massidda, F. Bernardini, C. Bersier, A. Continenza, P. Cudazzo, A. Floris, H. Glawe, M. Monni, S. Pittalis, G. Profeta, A. Sanna, S. Sharma, and E.~K.~U. Gross, Supercond. Sci. Technol. \textbf{22}, 034006 (2009).
\bibitem{footnote-diagrams} More detailed discussion about the correspondence of the kernels are seen in Ref.\onlinecite{Luders}
\bibitem{Gorling-Levy}A. G\"orling and M. Levy, Phys. Rev. A \textbf{50}, 196 (1994).
\bibitem{Hybertsen-Louie} M. S. Hybertsen and S. G. Louie, Phys. Rev. B \textbf{35}, 5585 (1987); M. S. Hybertsen and S. G. Louie, \textit{ibid}. \textbf{35}, 5602 (1987).
\bibitem{Pick-ASR} R. M. Pick, M. H. Cohen, and R. M. Martin, Phys. Rev. B \textbf{1}, 910 (1970).
\bibitem{Baroni-review} S. Baroni, S. de Gironcoli, A. Dal Corso, and P. Giannozzi, Rev. Mod. Phys. \textbf{73}, 515 (2001).
\bibitem{comment-kernel}For $\mathcal{K}^{\rm ph}_{n{\bm k},n'{\bm k}'}$ and $\mathcal{Z}^{\rm ph}_{n{\bm k}}$, we first calculate on logarithmic energy scales and next attribute each $\xi_{n{\bm k}}$ to the closest value. 
\bibitem{tetrahedra}G. Lehmann and M. Taut, Phys. Stat. Sol. \textbf{54}, 469 (1972). ; P. E. Blochl, O. Jepsen, and O. K. Andersen, Phys. Rev. B \textbf{49}, 16223 (1994).
\bibitem{Kawamura-priv} M. Kawamura, Y. Gohda, and S. Tsuneyuki, unpublished.
\bibitem{Lee-Chang-Nb} K. H. Lee and K. J. Chang, Phys. Rev. B \textbf{54}, 1419 (1996).
\bibitem{Ashcroft} N. W. Ashcroft and N. D. Mermin, \textit{Solid State Physics} (Thomson Learning, Singapore, 1976)
\bibitem{comment-approx} Another source of error is an assumption imposed in the derivation of Eq.~(\ref{eq:renorm}) cited from Ref.
\onlinecite{Morel-Anderson}, which corresponds to an assumption that the momentum average of the screened Coulomb interaction times density of states is constant. The drawback of this treatment is not clear.
\bibitem{QE} P. Giannozzi, S. Baroni, N. Bonini, M. Calandra, R. Car, C. Cavazzoni, D. Ceresoli, G. L. Chiarotti, M. Cococcioni, I. Dabo, A. Dal Corso, S. Fabris, G. Fratesi, S. de Gironcoli, R. Gebauer, U. Gerstmann, C. Gougoussis, A. Kokalj, M. Lazzeri, L. Martin-Samos, N. Marzari, F. Mauri, R. Mazzarello, S. Paolini, A. Pasquarello, L. Paulatto, C. Sbraccia, S. Scandolo, G. Sclauzero, A. P. Seitsonen, A. Smogunov, P. Umari, and R. M. Wentzcovitch, J. Phys.: Condens. Matter \textbf{21}, 395502 (2009); http://www.quantum-espresso.org/
\bibitem{PZ81}  J. P. Perdew and A. Zunger, Phys. Rev. B \textbf{23}, 5048 (1981).
\bibitem{Ceperley-Alder} D. M. Ceperley and B. J. Alder, Phys. Rev. Lett. \textbf{45}, 566 (1980).
\bibitem{Troullier-Martins} N. Troullier and J. L. Martins, Phys. Rev. B \textbf{43}, 1993 (1991).
\bibitem{Koelling-Harmons} D. D. Koelling and B. N. Harmons, J. Phys. C: Solid State Phys. \textbf{10}, 3107 (1977).
\bibitem{Louie-pcc}S. G. Louie, S. Froyen, and M. L. Cohen, Phys. Rev. B \textbf{26}, 1738 (1982).
\bibitem{MP-grid} H. J. Monkhorst and J. D. Pack, Phys. Rev. B \textbf{13}, 5188 (1976).
\bibitem{Methfessel-Paxton} M. Methfessel and A. T. Paxton, Phys. Rev. B \textbf{40}, 3616 (1989).
\bibitem{Shamoto-stacking} S. Shamoto, T. Kato, Y. Ono, Y. Miyazaki, K. Ohoyama, M. Ohashi, Y. Yamaguchi, and T. Kajitani, Physica C \textbf{306}, 7 (1998).
\bibitem{Adelmann} P. Adelmann, B. Renker, H. Schober, and F. Fernandez-Diaz, J. Low Temp. Phys. \textbf{117}, 449 (1999).
\bibitem{Chen-doped} X. Chen, L. Zhu, and S. Yamanaka, J. Solid State Chem. \textbf{169}, 149 (2002).
\bibitem{Chen-nondoped} X. Chen, T. Koiwasaki, and S. Yamanaka, J. Solid State Chem. \textbf{159}, 80 (2001).
\bibitem{Istomin} S. Istomin, J. K\"ohler, and A. Simon, Physica C \textbf{319}, 219 (1999).
\bibitem{Fuertes} A. Fuertes, M. Vlassov, D. Beltr\'an-Porter, P. Alemany, E. Canadell, N. Casa\~n-Pastor, and M. R. Palacin, Chem. Mater. \textbf{11}, 203 (1999).
\bibitem{Shamoto2004}  S. Shamoto, K. Takeuchi, S. Yamanaka, and T. Kajitani, Physica C \textbf{402}, 283 (2004).
\bibitem{Sugimoto-Oguchi} H. Sugimoto and T. Oguchi, J. Phys. Soc. Jpn. \textbf{73}, 2771 (2004).
\bibitem{Felser-band} C. Felser and R. Seshadri, J. Mater. Chem. \textbf{9}, 459 (1999).
\bibitem{comment-Wr} The detail of the interaction $W({\bf r, {\bf r}'})$ in a small $|{\bf r}-{\bf r}'|$ region, $|{\bf r}-{\bf r}'|$$\lesssim$$1$, does not affect an integrated value, because the integral weight scales as $O(|{\bf r}-{\bf r}'|^{2})$ while the interaction value itself scales as $O(|{\bf r}-{\bf r}'|^{-1})$.
\bibitem{Hotehama-molecular} K. Hotehama, T. Koiwasaki, K. Umemoto, S. Yamanaka, and H. Tou, J. Phys. Soc. Jpn. \textbf{79}, 014707 (2010).
\bibitem{Kawaji} H. Kawaji, K. Hotehama, and S. Yamanaka, Chem. Mater. \textbf{9}, 2127 (1997).
\bibitem{Kasahara-molecular} Y. Kasahara, T. Kishiume, K. Kobayashi, Y. Taguchi, and Y. Iwasa, Phys. Rev. B 82, 054504 (2010). 
\bibitem{Takano-optical} T. Takano, Y. Kasahara, T. Oguchi, I. Hase, Y. Taguchi, and Y. Iwasa, J. Phys. Soc. Jpn. \textbf{80}, 023702 (2011).
\bibitem{Cros} A. Cros, A. Cantarero, D. Beltr\'an-Porter, J. Or\'o-Sol\'e, and A. Fuertes, Phys. Rev. B \textbf{67}, 104502 (2003).
\end{thebibliography}
\end{document}